\documentclass[reqno,10pt,a4paper,table]{amsart}

\numberwithin{equation}{section}
\allowdisplaybreaks 

\usepackage{mathtools} 
\usepackage{amssymb,eucal,mathrsfs,setspace}
\usepackage[noadjust]{cite}
\usepackage{mathdots,tensor}
\usepackage{graphicx}
\usepackage{multirow}
\usepackage{slashbox}
\usepackage{verbatim}

\usepackage{caption} 

\usepackage{array}
\newcolumntype{C}{>{$}c<{$}}            
\newcommand{\RC}{\cellcolor[gray]{0.9}} 

\usepackage[largesc,theoremfont]{newtxtext}       
\usepackage[libertine,cmbraces,varbb]{newtxmath}  
\begingroup
  \makeatletter
  \@for\theoremstyle:=definition,remark,plain\do{%
    \expandafter\g@addto@macro\csname th@\theoremstyle\endcsname{%
      \addtolength\thm@preskip{.5\baselineskip plus .2\baselineskip minus .2\baselineskip}
      \addtolength\thm@postskip{.5\baselineskip plus .2\baselineskip minus .2\baselineskip}
    }%
  }
\endgroup

\usepackage[inner=25mm, outer=25mm, top=26mm, bottom=26mm, head=10mm, foot=10mm]{geometry}

\usepackage[colorlinks=true,citecolor=red,linkcolor=blue,urlcolor=blue]{hyperref} 

\usepackage{enumitem}
\setitemize{leftmargin=*}   
\setenumerate{leftmargin=*, 
  label=\textup{(\roman*)}} 

\usepackage[capitalise,noabbrev]{cleveref} 

\usepackage[dvipsnames]{xcolor}

\usepackage{tikz}
\usetikzlibrary{tikzmark}
\usetikzlibrary{calc}
\usepackage{pgfplots}
\usetikzlibrary{arrows,calc,decorations.markings}

\tikzset{nom/.style={circle,inner sep=2pt,fill=black!20!,minimum size=25pt}}

\newcommand{\pd}{\partial}     					

\renewcommand{\ge}{\geqslant} 
\renewcommand{\le}{\leqslant} 

\DeclareMathOperator{\id}{id}

\DeclareMathOperator{\End}{End}

\newcommand{\ii}{\mathfrak{i}} 
\newcommand{\cc}{\mathsf{c}}

\newcommand{\wun}{\vvmathbb{1}}

\DeclarePairedDelimiter{\brac}{\lparen}{\rparen} 		
\DeclarePairedDelimiter{\sqbrac}{\lbrack}{\rbrack} 		
\DeclarePairedDelimiter{\set}{\lbrace}{\rbrace}
\DeclarePairedDelimiter{\abs}{\lvert}{\rvert}

\DeclarePairedDelimiter{\normord}{{:}}{{:}}        				

\DeclarePairedDelimiterX{\comm}[2]{\lbrack}{\rbrack}{#1 , #2}  	
\DeclarePairedDelimiterX{\acomm}[2]{\lbrace}{\rbrace}{#1 , #2} 
\DeclarePairedDelimiterX{\inner}[2]{\langle}{\rangle}{#1 , #2} 	
\DeclarePairedDelimiterX{\super}[2]{\lparen}{\rparen}{#1 \delimsize\vert \mathopen{} #2} 

\newcommand{\lra}{\longrightarrow}
\newcommand{\lira}{\ensuremath{\lhook\joinrel\relbar\joinrel\rightarrow}}			 

\newcommand{\dses}[5]{0 \lra #1 \overset{#2}{\lra} #3 \overset{#4}{\lra} #5 \lra 0} 	    

\newcommand{\res}[1]{#1 \raisebox{0.05em}{$\downarrow$} {}} 
\newcommand{\ind}[1]{#1 \raisebox{0.08em}{$\uparrow$} {}}   

\newcommand{\fld}[1]{\mathbb{#1}}    		
\newcommand{\alg}[1]{\mathfrak{#1}} 	 	
\newcommand{\grp}[1]{\mathsf{#1}}    		
\newcommand{\Mod}[1]{\mathsf{#1}}    		
\newcommand{\VOA}[1]{\mathsf{#1}}    		
\newcommand{\categ}[1]{\mathscr{#1}} 		

\newcommand{\ZZ}{\fld{Z}}

\newcommand{\CC}{\fld{C}}

\newcommand{\SLG}[2]{\grp{#1} \brac*{#2}}       		 

\newcommand{\affine}[1]{\widehat{#1}}
\newcommand{\SLA}[2]{\alg{#1} \brac*{#2}}                 		
\newcommand{\AKMA}[2]{\affine{\alg{#1}} \brac*{#2}}       	  

\newcommand{\func}[2]{#1 \brac*{#2}}            	 
\newcommand{\parr}{\Pi}                      		 
\newcommand{\conjaut}{\gamma}                  	 
\newcommand{\sfaut}{\sigma}                      	 

\newcommand{\conjbcsymb}{\conjaut_{\supbc}}		 	                
\newcommand{\conjbc}[1]{\func{\conjaut_{\text{\supbc}}}{#1}}    
\newcommand{\sfbcsymb}{\sfaut_{\supbc}}		 		 	                
\newcommand{\sfbc}[2]{\func{\sfaut_{\text{\supbc}}^{#1}}{#2}}   

\newcommand{\conjscsymb}{\conjaut_{\supsc}}         		 
\newcommand{\sfscsymb}{\sfaut_{\supsc}}		 		 
\newcommand{\conjsc}[1]{\func{\conjscsymb}{#1}}                      
\newcommand{\sfsc}[2]{\func{\sfscsymb^{#1}}{#2}}         		 

\newcommand{\conjslsymb}{\conjaut_{\supsl}}		 		 	     
\newcommand{\sfslsymb}{\sfaut_{\supsl}}		 		 	        
\newcommand{\conjsl}[1]{\func{\conjslsymb}{#1}}          	 
\newcommand{\sfsl}[2]{\func{\sfslsymb^{#1}}{#2}}    		 

\newcommand{\conjfbsymb}{\conjaut_{\supH}}               
\newcommand{\sffbsymb}{\sfaut_{\supH}}                   
\newcommand{\conjfb}[1]{\func{\conjfbsymb}{#1}}          
\newcommand{\sffb}[2]{\func{\sffbsymb^{#1}}{#2}}         

\newcommand{\minmod}[2]{\VOA{M}\brac*{#1 , #2}}                    
\newcommand{\slminmod}[2]{\VOA{A}_1 \brac*{#1 , #2}}               
\newcommand{\bcghost}{\VOA{bc}}                                		
\newcommand{\fboson}{\VOA{H}}                                  		

\newcommand{\slcdimuni}[1]{\Delta_{#1}^{\supsl}} 
\newcommand{\slcdim}[2]{\Delta_{#1,#2}^{\supsl}} 
\newcommand{\sccdimuni}[2]{\Delta_{#1;#2}^{\supsc}} 
\newcommand{\sccdim}[3]{\Delta_{#1;#2,#3}^{\supsc}} 

\newcommand{\slirr}[1]{\Mod{L}_{#1}} 			
\newcommand{\sldis}[1]{\Mod{D}^+_{#1}}		
\newcommand{\sldism}[1]{\Mod{D}^-_{#1}}		
\newcommand{\sldispm}[1]{\Mod{D}^{\pm}_{#1}}
\newcommand{\sldismp}[1]{\Mod{D}^{\mp}_{#1}}
\newcommand{\slrel}[1]{\Mod{E}_{#1}} 			
\newcommand{\slrelp}[1]{\Mod{E}^+_{#1}} 		
\newcommand{\slrelpm}[1]{\Mod{E}^\pm_{#1}} 	
\newcommand{\slrelmp}[1]{\Mod{E}^\mp_{#1}} 	
\newcommand{\slproj}[1]{\Mod{S}_{#1}} 

\newcommand{\fock}[1]{\Mod{F}_{#1}}             

\newcommand{\bcmod}[1]{\Mod{N}_{#1}}
\newcommand{\bcnsp}{\bcmod{0}}                     
\newcommand{\bcnsm}{\bcmod{2}}                    
\newcommand{\bcrp}{\bcmod{1}}                        
\newcommand{\bcrm}{\bcmod{3}}                       

\newcommand{\NS}{\mathrm{NS}}
\newcommand{\R}{\mathrm{R}}

\newcommand{\Ver}[1]{\Mod{V}_{#1}}                             		
\newcommand{\NSVer}[3]{\Ver{#2 ; #3}^{\NS; #1}}     		
\newcommand{\RVer}[3]{\Ver{#2 ; #3}^{\R; #1}}          		
\newcommand{\Irr}[1]{\Mod{L}_{#1}}                		       		
\newcommand{\NSIrr}[3]{\Irr{#2 ; #3}^{\NS; #1}} 			
\newcommand{\RIrr}[3]{\Irr{#2 ; #3}^{\R; #1}}   			         
\newcommand{\NSRIrr}[3]{\Irr{#2 ; #3}^{\NS/\R; #1}} 			

\newcommand{\scirrrnu}[3]{{}^{[#1]}\Mod{C}_{#2;#3}}    	  		  
\newcommand{\scirrrn}[3]{{}^{[#1]}\Mod{C}^{\Mod{L}}_{#2;#3}}    	  

\newcommand{\scdisrn}[3]{{}^{[#1]}\Mod{C}^{\Mod{D}}_{#2;#3}}    	  

\newcommand{\screlrn}[3]{{}^{[#1]}\Mod{C}^{\Mod{E}}_{#2;#3}}    	  
\newcommand{\scarelrn}[3]{{}^{[#1]}\Mod{C}^{\pm}_{#2;#3}}    		  
\newcommand{\scarelrnp}[3]{{}^{[#1]}\Mod{C}^{+}_{#2;#3}}    		  

\newcommand{\scarelrnm}[3]{{}^{[#1]}\Mod{C}^{-}_{#2;#3}}    		  

\newcommand{\scproj}[3]{{}^{[#1]}\Mod{P}_{#2;#3}}                 

\DeclarePairedDelimiter{\kets}{\lvert}{\rangle}                                       						 
\DeclarePairedDelimiterX{\brakets}[2]{\langle}{\rangle}{#1 \delimsize\vert #2}                    			 
\DeclarePairedDelimiterX{\brackets}[3]{\langle}{\rangle}{#1 \delimsize\vert #2 \delimsize\vert #3}	 

\newcommand{\ket}[1]{\kets{#1}}

\DeclareMathOperator{\tr}{tr}

\newcommand{\traceover}[1]{\tr_{\raisebox{-2pt}{$\scriptstyle #1$}}}      
\newcommand{\chmap}{\mathrm{ch}}
\newcommand{\schmap}{\mathrm{sch}}
\newcommand{\Gr}[1]{\sqbrac[\big]{#1}}                                

\newcommand{\ch}[1]{\chmap \Gr{#1}}                                   
\newcommand{\fch}[2]{\ch{#1} \brac[\big]{#2}}                        
\newcommand{\sch}[1]{\schmap \Gr{#1}}                                
\newcommand{\fsch}[2]{\sch{#1} \brac[\big]{#2}}                     

\newcommand{\jth}[1]{\vartheta_{#1}}                                 
\newcommand{\fjth}[2]{\jth{#1} \brac{#2}}                            

\DeclareMathOperator*{\residue}{Res}
\newcommand{\resid}[2]{\residue_{#1}\sqbrac*{#2}}

\newcommand{\fuse}{\mathbin{\times}}                                      
\newcommand{\bfuse}{\mathbin{\dot{\times}}}                               
\newcommand{\Grfuse}{\mathbin{\boxtimes}}                                 

\newcommand{\vircoe}[3]{\mathrm{N}^{\mathrlap{(#1)}\phantom{#3} #2}_{#3}}					

\newcommand{\cft}{conformal field theory} 
\newcommand{\cfts}{conformal field theories}

\newcommand{\lcfts}{logarithmic conformal field theories}
\newcommand{\WZW}{Wess-Zumino-Witten}
\newcommand{\lw}{lowest-weight}

\newcommand{\hw}{highest-weight}
\newcommand{\hwv}{\hw{} vector}
\newcommand{\hwvs}{\hw{} vectors}
\newcommand{\hwm}{\hw{} module}
\newcommand{\hwms}{\hw{} modules}
\newcommand{\sv}{singular vector}
\newcommand{\svs}{\sv s}
\newcommand{\voa}{vertex operator algebra}
\newcommand{\voas}{\voa s}
\newcommand{\svoa}{vertex operator superalgebra}
\newcommand{\svoas}{\svoa s}
\newcommand{\ope}{operator product expansion}

\newcommand{\opes}{\ope s}
\newcommand{\lhs}{left-hand side}
\newcommand{\rhs}{right-hand side}

\newcommand{\rhss}{\rhs s}
\newcommand{\ns}{Neveu-Schwarz}

\newcommand{\AL}{\mathbf{AL}}

\theoremstyle{plain}

\newcommand{\commu}[2]{\operatorname{Com}(#2\,,#1)}   		
\newcommand{\algsl}{\AKMA{sl}{2}}                 				
\newcommand{\algbc}{\affine{\alg{bc}}}					          
\newcommand{\alghei}{\AKMA{gl}{1}}              					
\newcommand{\chvir}[1]{\chi^{\textup{Vir}}_{#1}}						
\newcommand{\fchvir}[1]{\chi^{\textup{Vir}}_{#1}(q)}					

\newcommand{\emt}{energy-momentum tensor}

\newcommand{\supH}{\textup{fb}}   
\newcommand{\supbc}{\textup{gh}}
\newcommand{\supsl}{\textup{aff}}
\newcommand{\supsc}{N=2}

\makeatletter
\renewcommand\author@andify{%
  \nxandlist {\unskip ,\penalty-1 \space\ignorespaces}%
    {\unskip {} \@@and~}%
    {\unskip \penalty-2 \space \@@and~}%
}
\makeatother

\begin{document}

\title{Unitary and non-unitary $N=2$ minimal models}

\author[T~Creutzig]{Thomas Creutzig}
\address[Thomas Creutzig]{
Department of Mathematical and Statistical Sciences \\
University of Alberta \\
Edmonton, Canada T6G~2G1 and \\
Research Institute for Mathematical Sciences \\ Kyoto University\\ Kyoto Japan 606-8502.
}
\email{creutzig@ualberta.ca}

\author[T~Liu]{Tianshu Liu}
\address[Tianshu Liu]{
School of Mathematics and Statistics \\
University of Melbourne \\
Parkville, Australia, 3010.
}
\email{tianshul@student.unimelb.edu.au}

\author[D~Ridout]{David Ridout}
\address[David Ridout]{
School of Mathematics and Statistics \\
University of Melbourne \\
Parkville, Australia, 3010.
}
\email{david.ridout@unimelb.edu.au}

\author[S~Wood]{Simon Wood}
\address[Simon Wood]{
School of Mathematics \\
Cardiff University \\
Cardiff, United Kingdom, CF24 4AG.
}
\email{woodsi@cardiff.ac.uk}

\begin{abstract}
	The unitary $N=2$ superconformal minimal models have a long history in string theory and mathematical physics, while their non-unitary (and logarithmic) cousins have recently attracted interest from mathematicians.  Here, we give an efficient and uniform analysis of all these models as an application of a type of Schur-Weyl duality, as it pertains to the well-known Kazama-Suzuki coset construction.  The results include straightforward classifications of the irreducible modules, branching rules, (super)characters and (Grothendieck) fusion rules.
\end{abstract}

\maketitle

\onehalfspacing

\section{Introduction}

\subsection{Background}

$N=2$ supersymmetry is ubiquitous in string theory where its first appearances even predate the conception of \cft{} as a separate discipline, see \cite{AdeSup76} for example.  Upon formalising conformal invariance, physicists quickly started exploring the properties of the $N=2$ superconformal algebra \cite{DiVN=285,BouDet86,NamKac86,SchCom87} and its representations, especially the unitary ones \cite{ZamDis86,DiVUni86,BouDet86,DiVExp86,EguUni88,LerChi89}.  The discovery \cite{KazNew89,KazCha89} of a coset construction for the corresponding minimal models led to many generalisations, now known as Kazama-Suzuki models, and important links to the geometry of string compactifications.

On the representation-theoretic side, the unitary $N=2$ superconformal minimal models were studied by mathematicians and physicists interested in their characters \cite{DobCha87,MatCha87,DorSin95,EhoUni96,DorEmb98}, modularity \cite{RavMod87,QiuMod87} and fusion rules \cite{WakFus98,ADAMOVIC01}.  Their non-unitary cousins unfortunately attracted relatively little attention, though a new construction as a minimal quantum hamiltonian reduction \cite{KacQua03,KacQua04} realised an important link with mock modular forms \cite{SemHig05,KacRep14,KacRep16,KacRep17}.  Moreover, their Kazama-Suzuki coset relationship with the fractional-level $\SLA{sl}{2}$ \WZW{} models was reformulated into a number of beautiful categorical equivalences \cite{FeiEqu98,SemEmb97,FeiRes98,AdaRep99,SatEqu16,SatMod17,Koshida:2018lxt}.

With these fractional-level models now well in hand \cite{AdaVer95,GabFus01,RidSL208,RidSL210,RidFus10,CreMod12,CreMod13,RidRel15}, this relationship can be exploited in both directions.  Our aim here is to use this knowledge to give a uniform and direct treatment of the $N=2$ superconformal minimal models, both unitary and non-unitary, with the main results being a classification of irreducible modules, explicit branching rules and characters, and (Grothendieck) fusion rules.  The point is that we have established an efficient procedure to extract representation theory from coset constructions: the $N=2$ superconformal minimal models provide a beautiful and important illustration of these methods.

\subsection{A Schur-Weyl duality for Heisenberg cosets}

Over the last few years, in a joint effort with Shashank Kanade, Robert McRae and Andrew Linshaw, two of the authors have developed a working theory of coset \voas{} \cite{CreCos14,Cre2015ar,CreSch16,CreTen17}.  This has been strongly influenced by physics ideas, but builds on the work of many mathematicians including Kac--Radul \cite{KacRep96}, Dong--Li--Mason \cite{DonCom96}, Huang--Lepowsky--Zhang \cite{HuaLog10} and Huang-Kirillov-Lepowsky \cite{HKL}.  The present paper is one of a series that applies this new technology to interesting examples.

The picture is the following. We have a vertex operator (super)algebra $\VOA{V}$ that contains two mutually commuting subalgebras $\VOA{A}$ and $\VOA{C}$.  Assuming that we understand the (relevant) representation theories of $\VOA{A}$ and $\VOA{V}$, we aim to extract the representation theory of the coset algebra $\VOA{C}$.  This works particularly well if $\VOA{A}$ is a Heisenberg \voa{} (acting diagonalisably on $\VOA{V}$).  Then, we are precisely in the situation of \cite{CreSch16} in which we have established a Schur-Weyl-type duality between $\VOA{C}$- and $\VOA{V}$-modules.  The branching rules, which indicate how any given $\VOA{V}$-module decomposes into a direct sum of Fock spaces tensored with coset modules, are thereby known to be structure-preserving: each $\VOA{V}$-module begets an infinite number of $\VOA{C}$-modules, each labelled by a Fock space weight (momentum), whose structures (Loewy diagrams and radical/socle series) are equivalent to that of the parent $\VOA{V}$-module \cite[Thm.~3.8]{CreSch16}.  In particular, each irreducible $\VOA{V}$-module yields an infinite number of irreducible $\VOA{C}$-modules.  Moreover, every indecomposable $\VOA{C}$-module (under some mild conditions) may be tensored with a Fock space so that the product lifts to a $\VOA{V}$-module \cite[Thm.~4.3]{CreSch16}.

This lifting procedure is mathematically implemented by an induction functor.  Happily, this functor is monoidal \cite{CreTen17}, meaning that the fusion product of two induced $\VOA{C}$-modules, which are $\VOA{V}$-modules, is isomorphic to the result of fusing the $\VOA{C}$-modules and then inducing \cite{RidVer14}.  It follows that one can determine the fusion rules of $\VOA{C}$ if those of $\VOA{V}$ are known, and vice versa.  We have already applied this powerful realisation to the example of non-unitary (logarithmic) parafermions in \cite{AugMod17}.  A similar application involving a non-Heisenberg coset (and the vice versa direction) has also recently appeared \cite{CreRep17,CreCos18}.  The example that concerns us here has $\VOA{V}$ as the tensor product of the simple affine \voa{} of $\SLA{sl}{2}$, at admissible level $k=-2+\frac{u}{v}$, and the fermionic ghost \svoa{} (of central charge $1$).  Here, $u$ and $v$ are coprime positive integers with $u>1$.  We recall that the $N=2$ minimal models and the fractional-level $\SLA{sl}{2}$ models are only unitary when $v=1$.

\subsection{Characters and meromorphic Jacobi forms}

There are of course subtleties to overcome when dealing with the non-unitary $N=2$ minimal models ($v>1$).  In this case, we are guided by the standard module formalism \cite{CreLog13,RidVer14} that has worked so well in analysing similar \lcfts{}.  In particular, it applies \cite{CreMod12,CreMod13} to the fractional-level $\SLA{sl}{2}$ \WZW{} models that appear in the (non-unitary) $N=2$ coset construction.  In this case, the characters of the standard modules \cite{AdaRea17,KawRel18} are naturally expressed as distributions in the Jacobi variable that keeps track of the Cartan weight.  They have exemplary modular properties and the standard Verlinde formula gives non-negative fusion multiplicities.  However, there are other ``atypical'' modules whose characters naturally extend \cite{KacMod88} to meromorphic Jacobi forms of index $k$ (the forms are only holomorphic if $v=1$).  The modularity of these forms is somewhat infamous: a na\"{\i}ve application of the Verlinde formula results in negative multiplicities \cite{KohFus88}.  The standard module formalism (correctly) rejects these meromorphic extensions and instead regards the atypical characters as infinite linear combinations of standard ones.  This formally resolves the negative multiplicity issue, but these infinite linear combinations turn out to diverge when $k>0$.

Similar divergences also plague the atypical characters of the non-unitary $N=2$ minimal models when we apply the methods of the standard module formalism.  Indeed, we shall explicitly demonstrate below that these characters converge for $k<0$ and diverge otherwise, when treated as distributions.  However, the story differs markedly from that of the $\SLA{sl}{2}$ models in that the $N=2$ atypical modules decompose into finite-dimensional eigenspaces under the action of the Virasoro zero mode.  Their characters must therefore converge for all non-zero values of their Jacobi variable, hence we must have convergence as functions for all $k$.  To take advantage of this, we therefore need to rethink our character methods.

Going back to the meromorphic Jacobi forms of the $\SLA{sl}{2}$ models, we recall that obtaining their Fourier decompositions, the character analogue of coset branching rules, is generally considered rather difficult.  However, it may be solved \cite{Dabholkar:2012nd,Bringmann2014,Bringmann2016} by computing some very delicate contour integrals.  Interestingly, the resulting Fourier coefficients turn out to be mock modular forms in general.  However, we expect that these computations would be quite cumbersome in our situation.

In \cite{AugMod17}, the logarithmic parafermion algebras of $\SLA{sl}{2}$, with $k<0$, were studied, along with their infinite-order simple current extensions (which are expected to be $C_2$-cofinite).  There, the modularity of these extensions was analysed without resorting to contour integral machinery, despite having to deal with (negative-index) meromorphic Jacobi forms.  Inspired by this, we have found a way to uniformly deal with the (negative- and positive-index) meromorphic Jacobi forms that arise in the $N=2$ coset.  The key is a ``magic identity'' \cite[Eqs.~(A.3--4)]{CreMod12} that has already played an important role in studying the modularity of the fractional-level $\SLA{sl}{2}$ models.  Here, we employ it once again to straightforwardly Fourier-decompose the meromorphic Jacobi forms that arise and so deduce convergent character formulae for the atypical modules of the non-unitary $N=2$ minimal models.  They turn out to be expressible in terms of higher-level Appell-Lerch sums \cite{SemHig05}, see also \cite{KacRep16,SatMod17}.

We emphasise that the resulting atypical $N=2$ characters are holomorphic in the Jacobi variable (on the punctured plane), as required.  It is therefore reasonable to suppose that they have excellent (mock) modular properties and, in particular, that applying the standard Verlinde formula will result in non-negative fusion multiplicities.  We shall not attempt to confirm this here because, as noted above, we are able to attack the problem of determining the fusion rules directly using induction.  However, we note that this supposition is encouraged by the example recently computed by Sato \cite[Ex.~5.2]{SatMod17}, see also \cite[Rem.~5.14]{Koshida:2018lxt}, who indeed finds non-negative multiplicities for one particular fusion rule of the $N=2$ minimal model of central charge $-1$ ($u=3$, $v=2$).  To the best of our knowledge, this is the first (and only) Verlinde calculation that has been performed for a non-unitary $N=2$ minimal model.  We hope that this will shortly change.

\subsection{Outline and Results}

We begin, in \cref{sec:N=2}, with a discussion of the $N=2$ \svoas{} and their associated \ns{} and Ramond Lie superalgebras.  Their \hw{} representation theories are reviewed with particular emphasis on spectral flow automorphisms \cite{SchCom87} and the action of these automorphisms on $N=2$ modules.  The ingredients of the coset construction, being the free boson, fermionic ghosts and the fractional-level $\SLA{sl}{2}$ \WZW{} models, are introduced in \cref{sec:coset} in order to fix notation and review those aspects that will be crucial for what follows.  In \cref{embed}, we explicitly embed the (simple!) $N=2$ algebra into the tensor product of the corresponding fractional-level $\SLA{sl}{2}$ model and a fermionic ghost system, thereby giving a very quick proof of this instance of the Kazama-Suzuki coset construction.

\cref{sec:branch} is devoted to explicitly computing the branching rules of the coset.  This amounts to decomposing tensor products of affine and ghost modules into Fock spaces and $N=2$ modules.  By fully exploiting the spectral flow automorphisms of all of the vertex algebras involved, the calculations are efficiently reduced to a small number that are easily dealt with using the method of ``extremal states'' \cite{FeiEqu98}.  We then build dictionaries, for the unitary and non-unitary models in both the \ns{} and Ramond sectors, that identify the modules appearing in the branching rules as modules over the appropriate $N=2$ Lie superalgebra (including their global parities).  We also prove that the irreducible $N=2$ modules that arise in these branching rules exhaust all the irreducible weight modules of the $N=2$ \svoa{}.

The characters of these $N=2$ modules are then studied in \cref{sec:char}.  The basic tool used is the same as that used by Eholzer--Gaberdiel to compute specialised characters for the unitary models \cite{EhoUni96} --- we extend their method to obtain the full characters, again for the unitary models.  Unfortunately, technical issues restrict us to considering only a small subset of the characters.  However, it turns out that this subset includes members of every spectral flow orbit, allowing us to obtain a complete set of full characters (and supercharacters).

We then turn to the (full) characters of the non-unitary models using the same method.  As previewed above, this works perfectly for the standard modules but encounters divergence issues for the atypical modules with $k>0$ when the ``resolution'' formulae for the $\SLA{sl}{2}$ models \cite{CreMod13} are used.  In the atypical case, we instead apply a ``magic identity'' from \cite{CreMod12} to the admissible Kac--Wakimoto characters \cite{KacMod88} which allows us to extract convergent $N=2$ characters.  The atypical characters, along with the unitary characters, may be succinctly represented using higher-level Appell-Lerch sums \cite{SemHig05}.  We remark that the ``magic identity'' actually recovers the unitary characters more efficiently than the method of Eholzer--Gaberdiel (and without having to resort to spectral flow).

\cref{sec:fusion} addresses the fusion rules of the unitary and non-unitary $N=2$ minimal models, employing an induction functor as the main tool.  Inducing from Fock spaces tensored with $N=2$ modules to affine and ghost modules allows us to compute the fusion rules of the former in terms of those of the latter.  We illustrate this simple method with the unitary models whose fusion rules are completely determined.  For the non-unitary models, we can only compute some of the fusion rules (because those of the $\SLA{sl}{2}$ models are not known in general).  However, we do obtain all of their Grothendieck fusion rules and make some conjectures regarding some of the remaining $N=2$ fusion rules.  These conjectures involve some staggered $N=2$ modules, these being indecomposable with a non-diagonalisable action of the Virasoro zero mode (this indicates the logarithmic nature of the non-unitary $N=2$ minimal models).  We conclude by conjecturing that these staggered modules are actually projective in an appropriate category of $N=2$ models.

\section*{Acknowledgements}

We thank Chris Raymond for a thorough proof-reading and helpful comments.
TC is supported by the Natural Sciences and Engineering Research Council of Canada (RES0020460).
TL's research is supported by a University Research Scholarship from the University of Melbourne.
DR's research is supported by the Australian Research Council Discovery Project DP160101520 and the Australian Research Council Centre of Excellence for Mathematical and Statistical Frontiers CE140100049.
SW's research is supported by the Australian Research Council Discovery Early Career Researcher Award DE140101825 and the Discovery Project DP160101520.

\section{$N=2$ algebras} \label{sec:N=2}

\subsection{Algebraic preliminaries} \label{sec:algprelim}

The $N=2$ superconformal algebras, or $N=2$ algebras for short, are a family of \svoas{} parametrised by their central charges $\cc \in \CC$ with each being strongly generated by two bosonic fields $J(z)$ and $T^{\supsc}(z)$ and two fermionic fields $G^+(z)$ and $G^-(z)$.  Here, $T^{\supsc}$ is the \emt{} and $J$, $G^+$ and $G^-$ are Virasoro primaries of conformal dimensions $1$, $\frac{3}{2}$ and $\frac{3}{2}$, respectively.  The \opes{} between these fields are given by
\begin{equation} \label{ope:N=2}
	\begin{gathered}
		J(z) J(w) \sim \frac{(\cc/3) \wun}{(z-w)^2}, \qquad
		J(z) G^{\pm}(w) \sim \frac{\pm G^{\pm}(w)}{z-w}, \qquad
		G^{\pm}(z) G^{\pm}(w) \sim 0, \\
		G^{\pm}(z) G^{\mp}(w) \sim \frac{(2\cc/3) \wun}{(z-w)^3} \pm \frac{2\,J(w)}{(z-w)^2} + \frac{2\,T^{\supsc}(w) \pm \pd J(w)}{z-w},
	\end{gathered}
\end{equation}
where $\wun$ is the identity field. We shall distinguish between the \emph{universal} $N=2$ algebra of central charge $\cc$, in which the \opes{} \eqref{ope:N=2} generate a complete set of relations, and the \emph{minimal model} $N=2$ algebras.  The latter are only defined when the universal $N=2$ algebra is not simple.  This occurs if and only if \cite{GorSim07}
\begin{equation} \label{eq:N=2MinModc}
	\cc = 3 - \frac{6v}{u}, \qquad u \in \ZZ_{\ge 2},\ v \in \ZZ_{\ge 1},\ \gcd \set{u,v} = 1.
\end{equation}
In this case, the minimal model is defined to be the unique simple quotient of the universal $N=2$ algebra and will be denoted by $\minmod{u}{v}$.

The boundary conditions imposed on the fermionic fields determines their Fourier mode expansions.
This gives rise to three sectors in the representation theory of the $N=2$ algebra.
\begin{center}
	\begin{tabular}{l|CCCC}
		& L^{\supsc}_m & J_n & G^+_r & G^-_s \\
		\hline
		\ns{} & m \in \ZZ & n \in \ZZ & r \in \ZZ + \frac{1}{2} & s \in \ZZ + \frac{1}{2} \\
		Ramond & m \in \ZZ & n \in \ZZ & r \in \ZZ & s \in \ZZ \\
		Twisted & m \in \ZZ & n \in \ZZ + \frac{1}{2} &- & -
	\end{tabular}
\end{center}
The moding in the twisted sector is not well adapted to the $G^{\pm}_n$ basis elements. This sector is included for completion only and will not be studied in the rest of the paper.

The mutual localities of the generating fields follows standard boson-fermion statistics.  In terms of Lie brackets, the non-vanishing (anti-)commutation relations between the modes are thus
\begin{equation} \label{eq:N=2Comm}
	\begin{gathered}
		\comm{L^{\supsc}_m}{L^{\supsc}_n} = (m-n) L^{\supsc}_{m+n} + \tfrac{1}{12} (m^3-m) \delta_{m+n,0} \cc \wun, \\
		\begin{aligned}
			\comm{L^{\supsc}_m}{J_n} &= -n J_{m+n}, &&&&&&& \comm{J_m}{J_n} &= \tfrac{1}{3} m \delta_{m+n,0} \cc \wun, \\
			\comm{L^{\supsc}_m}{G^{\pm}_s} &= (\tfrac{1}{2} m - s) G^{\pm}_{m+s}, &&&&&&& \comm{J_m}{G^{\pm}_s} &= \pm G^{\pm}_{m+s},
		\end{aligned}
		\\
		\acomm{G^{\pm}_r}{G^{\mp}_s} = 2 L^{\supsc}_{r+s} \pm (r-s) J_{r+s} + \tfrac{1}{12} (4r^2-1) \delta_{r+s,0} \cc \wun,
	\end{gathered}
\end{equation}
where the mode indices are sector-dependent, as above, leading to a \ns{} and a Ramond $N=2$ Lie superalgebra for each value of the central charge $\cc$.  Here, $\wun$ should be interpreted as a central element of the superalgebra.

The $N=2$ Lie superalgebras admit many automorphisms including the \emph{conjugation} automorphism $\conjscsymb$, given by
\begin{subequations} \label{eq:N=2auts}
	\begin{equation} \label{auto2}
		\conjsc{L^{\supsc}_n} = L^{\supsc}_n, \quad \conjsc{J_n} = -J_n, \quad \conjsc{G^{\pm}_s} = G^{\mp}_s,\quad \conjsc{\wun}=\wun,
	\end{equation}
		and the \emph{spectral flow} automorphisms $\sfscsymb^{\ell}$, $\ell \in \ZZ$, given by
	\begin{equation} \label{sfgenerator}
		\sfsc{\ell}{L^{\supsc}_n} = L^{\supsc}_n - \ell J_n + \tfrac{1}{6} \ell^2 \delta_{n,0} \cc \wun, \quad
		\sfsc{\ell}{J_n} = J_n - \tfrac{1}{3} \ell \delta_{n,0} \cc \wun, \quad
		\sfsc{\ell}{G^{\pm}_s} = G^{\pm}_{s \mp \ell},\quad
		\sfsc{\ell}{\wun}=\wun.
	\end{equation}
\end{subequations}
Taking $\ell \in \ZZ + \frac{1}{2}$ changes the moding of the fermions, meaning that half-integer spectral flows define isomorphisms between the \ns{} and Ramond $N=2$ Lie superalgebras. Both conjugation and spectral flow lift to automorphisms of the universal $N=2$ vertex superalgebras as well as their minimal model quotients. However, spectral flow does not define automorphisms of the $N=2$ superconformal \svoas{} because it does not preserve the \emt{}. Note that because \(\conjscsymb^2=\id\), \(\sfscsymb^\ell \conjscsymb = \conjscsymb \sfscsymb^{-\ell}\) and \(\sfscsymb\) has infinite order, the group generated by \(\conjscsymb\) and \(\sfscsymb^{1/2}\) is isomorphic to the infinite dihedral group.

\subsection{Highest-weight representations} \label{sec:N=2Reps}

Consider the subalgebra of the $N=2$ \ns{} Lie superalgebra of central charge $\cc \in \CC$ that is spanned by $\wun$ and the modes with non-negative indices.  For $j,\Delta \in \CC$, let $\CC^{\NS;+}_{j;\Delta}$ denote the one-dimensional module of this subalgebra that is spanned by a bosonic state on which every mode acts as zero, except for $J_0$, $L^{\supsc}_0$ and $\wun$ which act as multiplication by $j$, $\Delta$ and $1$, respectively. Inducing this to a module over the full $N=2$ \ns{} Lie superalgebra now gives the $N=2$ \ns{} Verma module $\NSVer{+}{j}{\Delta}$. If we alter $\CC^{\NS;+}_{j;\Delta}$ so that the spanning state is fermionic, then the resulting Verma module will be denoted by $\NSVer{-}{j}{\Delta}$.

We shall always assume that the modules of a superalgebra, such as the $N=2$ Lie superalgebras and their associated \svoas{}, are $\ZZ_2$-graded, meaning that they decompose into a direct sum of two subspaces, called the bosonic and fermionic subspaces, which are preserved by the action of the bosonic elements and are swapped by the action of the fermionic elements.  In general, we have the \emph{parity reversal functor} $\parr$ which redefines all bosonic states to be fermionic and all fermionic ones to be bosonic.  For example, $\parr \NSVer{+}{j}{\Delta} \cong \NSVer{-}{j}{\Delta}$.

As usual, \ns{} Verma modules have unique maximal proper submodules and we shall denote their irreducible quotients by $\NSIrr{\pm}{j}{\Delta}$. Both $\NSVer{\pm}{j}{\Delta}$ and $\NSIrr{\pm}{j}{\Delta}$ are modules over the universal $N=2$ algebra of central charge $\cc$ (which is left implicit). A \ns{} \hwv{} is a simultaneous eigenvector of $J_0$, $L^{\supsc}_0$ and $\wun$ that is annihilated by every mode of positive index.  We shall say that a \sv{} of a given module is a \hwv{} that does not generate the entire module.  For example, if $v$ denotes the generating \hwv{} of $\NSVer{+}{0}{0}$, then $G^+_{-1/2} v$ and $G^-_{-1/2} v$ are both \svs{}.  Quotienting by the sum of the submodules that they generate results in the vacuum module of the universal $N=2$ algebra.

In the Ramond sector, one defines Verma modules by choosing a triangular decomposition such that $G^+_0$ is an annihilation operator and $G^-_0$ is a creation operator.  For $j, \Delta \in \CC$, let $\CC_{j;\Delta}^{\R; \pm}$ be the one-dimensional module of even ($+$) or odd ($-$) parity over the Ramond subalgebra (of central charge $\cc$) spanned by $\wun$, $J_0$, $L^{\supsc}_0$, $G^+_0$ and the positive index modes, where every mode acts as zero except $J_0$, $L^{\supsc}_0$ and $\wun$ which act as multiplication by $j$, $\Delta$ and $1$, respectively.  Inducing then gives the $N=2$ Ramond Verma module $\RVer{\pm}{j}{\Delta}$.  It, and its irreducible quotient $\RIrr{\pm}{j}{\Delta}$, are ($\ZZ_2$-twisted) modules over the universal $N=2$ algebra of central charge $\cc$.

A Ramond \hwv{} is then a simultaneous eigenvector of $J_0$, $L^{\supsc}_0$ and $\wun$ that is annihilated by all the modes of positive index and $G^+_0$, while a Ramond \sv{} is a Ramond \hwv{} that does not generate the entire module.  Let $v$ be a Ramond \hwv{} of charge ($J_0$-eigenvalue) $j$ and conformal dimension ($L^{\supsc}_0$-eigenvalue) $\Delta$.  It satisfies
\begin{equation}
	G^+_0 G^-_0 v = 2 (\Delta - \tfrac{\cc}{24}) v.
\end{equation}
When $\Delta = \frac{\cc}{24}$, $G^-_0 v$ is thus a \sv{}, so $\RIrr{\pm}{j}{\Delta}$ has a one-dimensional space of ground states spanned by $v$. When $\Delta \neq \frac{\cc}{24}$, $G^-_0 v$ is not singular and $\RIrr{\pm}{j}{\Delta}$ has a two-dimensional space of ground states spanned by $v$ and $G^-_0 v$.  Note that these states have charges $j$ and $j-1$, respectively; their common conformal dimension is $\Delta$.

It is always useful to consider families of modules that are related by twisting by an automorphism $\omega$.  As we want to distinguish between the elements of the module $\Mod{M}$ and those of the resulting twisted module, we let $\omega^*$ denote an (arbitrary) vector space isomorphism from $\Mod{M}$ to the twisted module, hereafter denoted by $\omega^*(\Mod{M})$, equipping the latter with the following algebra action:
\begin{equation} \label{auto1}
	x \cdot \omega^*(m) = \omega^*(\omega^{-1}(x) \cdot m), \quad \text{for all modes \(x\) and \(m \in \Mod{M}\).}
\end{equation}
This action promotes $\omega^*$ to an invertible (and therefore structure-preserving) functor on an appropriate module category.  The categories of interest here are the \ns{} and Ramond weight modules over either the universal or minimal model $N=2$ algebra, for fixed central charge.  In what follows, we will generally drop the star that distinguishes an automorphism from the corresponding functor.

Twisting the action on an $N=2$ module $\Mod{M}$, by acting with the automorphisms $\conjscsymb$ or $\sfscsymb^{\ell}$ from \eqref{eq:N=2auts}, results in modules $\conjsc{\Mod{M}}$ and $\sfsc{\ell}{\Mod{M}}$.  We shall refer to these modules as the conjugate and the spectral flow of $\Mod{M}$, respectively. To illustrate this, suppose that $v$ is a weight vector of charge $j$ and conformal dimension $\Delta$.  Then, it is easy to check using \eqref{eq:N=2auts} and \eqref{auto1} that $\conjsc{v}$ and $\sfsc{\ell}{v}$ are weight vectors satisfying
\begin{equation} \label{eq:SFWts}
	\begin{aligned}
		L^{\supsc}_0 \conjsc{v} &= \Delta\,\conjsc{v}, & J_0 \conjsc{v} &= -j\,\conjsc{v}, \\
		L^{\supsc}_0 \sfsc{\ell}{v} &= (\Delta + \ell j + \tfrac{1}{6} \ell^2 \cc) \,\sfsc{\ell}{v}, & J_0\sfsc{\ell}{v} &= (j + \tfrac{1}{3} \ell \cc)\,\sfsc{\ell}{v}.
	\end{aligned}
\end{equation}
From this, we deduce the following isomorphisms among irreducible $N=2$ modules:
\begin{equation}
	\begin{aligned}
		\conjsc{\NSIrr{\pm}{j}{\Delta}} &\cong \NSIrr{\pm}{-j}{\Delta}, &
		\sfsc{1/2}{\NSIrr{\pm}{j}{\Delta}} &\cong \RIrr{\pm}{j+\cc/6}{\Delta+j/2+\cc/24}, \\
		\conjsc{\RIrr{\pm}{j}{\Delta}} &\cong
		\begin{cases*}
			\RIrr{\pm}{-j}{\Delta}, & if $\Delta = \frac{\cc}{24}$, \\
			\RIrr{\mp}{-j+1}{\Delta}, & otherwise,
		\end{cases*}
		& \sfsc{1/2}{\RIrr{\pm}{j}{\Delta}} &\cong
		\begin{cases*}
			\NSIrr{\pm}{j+\cc/6}{(j+\cc/6)/2}, & if $\Delta = \frac{\cc}{24}$, \\
			\NSIrr{\mp}{j-1+\cc/6}{\Delta+(j-1)/2+\cc/24}, & otherwise.
		\end{cases*}
	\end{aligned}\label{sfmod}
\end{equation}

\section{The coset construction} \label{sec:coset}

Recall that $\minmod{u}{v}$ denotes the $N=2$ minimal model of central charge $\cc$, given in \eqref{eq:N=2MinModc}.  As is well known, see \cite{EhoUni96} for an early reference and Lemma 8.6 of \cite{CreCos14} for a proof, this minimal model may be represented as the following coset (commutant):
\begin{equation} \label{cosetN=2}
	\minmod{u}{v} =\commu{\slminmod{u}{v} \otimes \bcghost}{\fboson} = \frac{\slminmod{u}{v} \otimes \bcghost}{\fboson}.
\end{equation}
Here, $\slminmod{u}{v}$, $\bcghost$ and $\fboson$ denote the simple \svoas{} associated to the affine algebra $\algsl$ at level $k=-2+\frac{u}{v}$, the fermionic ghost algebra $\algbc$ and the Heisenberg (free boson) algebra $\alghei$, respectively.  We begin with a detailed discussion of these three component superalgebras.  We then describe the embedding
\begin{equation} \label{eq:embedding}
	\fboson \otimes \minmod{u}{v} \lira \slminmod{u}{v} \otimes \bcghost
\end{equation}
in detail in order to facilitate the later analysis.

\subsection{The Heisenberg algebra} \label{subsec:fb}

The \voa{} $\fboson$ associated to the Heisenberg algebra $\alghei$ is generated by a single bosonic field $a(z)$, whose \ope{} with itself is given by
\begin{equation} \label{fbOPE}
	a(z)a(w)\sim\frac{2t\wun}{(z-w)^2}.
\end{equation}
Here, we have scaled the \rhs{} by $2t$ with respect to the usual conventions in the literature, where $t \in \CC \setminus \set{0}$, for later convenience. The modes of the generating field therefore satisfy the $\alghei$ commutation relations
\begin{equation}
	\comm{a_m}{a_n} = 2t m\delta_{m+n,0} \wun, \quad m,n\in\ZZ,
\end{equation}
and we choose the \emt{} to be
\begin{equation}
	T^{\supH}(z)=\frac 1 {4t} \normord{aa}(z).
\end{equation}
With this choice, the central charge is $1$ and $a(z)$ is primary of conformal dimension $1$.  The label reflects the fact that the corresponding \cft{} describes a free boson (in a single spacetime dimension).

The \hwms{} of $\alghei$, called Fock spaces (of charge $p \in \CC$) and denoted by $\fock{p}$, are Verma modules induced in the same way as those of the $N=2$ Lie superalgebras (\cref{sec:N=2Reps}).  We first define a one-dimensional module $\CC_{p}$ on which the annihilators (the $a_n$ with $n\ge1$) act trivially while $a_0$ and $\wun$ act as multiplication by $p$ and $1$, respectively. We then induce $\CC_{p}$ to a $\alghei$-module $\fock{p}$ by having the creators (the $a_{-n}$ with $n\ge1$) act freely. The resulting Fock space is irreducible.  Its character is given by
\begin{equation} \label{charF}
	\fch{\fock{p}}{y;q}=\traceover{\fock{p}}y^{a_0}q^{L^{\supH}_0-1/24} = \frac{y^p q^{p^2/4t}}{\eta(q)},
\end{equation}
where $\eta(q)=q^{1/24}\prod_{i=1}^{\infty}(1-q^i)$ is Dedekind's eta function and the $L^{\supH}_n$ are the modes of $T^{\supH}(z)$.  The fusion rules of the Fock spaces are well known:
\begin{equation} \label{eq:frH}
	\fock{p} \fuse \fock{p'} \cong \fock{p+p'}.
\end{equation}
The vacuum module, meaning the one that carries the structure of the \voa{} $\fboson$, is $\fock{0}$.

The Lie algebra $\alghei$ admits a conjugation automorphism $\conjfbsymb$ and spectral flow automorphisms $\sffbsymb^{p'}$, $p' \in \CC$, given by
\begin{subequations} \label{eq:fbauts}
	\begin{align}
		\conjfb{a_n} &= -a_n, & \conjfb{L^{\supH}_n} &= L^{\supH}_n, & \conjfb{\wun} &= \wun; \\
		\sffb{p'}{a_n} &= a_n - p' \delta_{n,0} \wun, & \sffb{p'}{L^{\supH}_n} &= L^{\supH}_n - p' a_n + \tfrac{1}{2} {p'}^2 \delta_{n,0} \wun, & \sffb{p'}{\wun} &= \wun.
	\end{align}
\end{subequations}
These automorphisms generate a generalised dihedral group of \(\CC\) (with addition as the binary operation).  They also lift to automorphisms of the Heisenberg vertex algebra.  It is easy to check that the induced action on the Fock spaces is given by
\begin{equation}\label{eq:Hauto}
	\conjfb{\fock{p}} \cong \fock{-p}, \qquad \sffb{p'}{\fock{p}} \cong \fock{p+p'},
\end{equation}
which neatly explains the identical structures of the Fock spaces (they are all simple).

\subsection{The fermionic ghost algebra}\label{subsec:bc}

The ghost \svoa{} $\bcghost$ is generated by two fermionic fields, denoted by $b(z)$ and $c(z)$, which satisfy
\begin{equation}
	b(z)c(w)\sim c(z)b(w)\sim\frac{\wun}{z-w}, \qquad b(z)b(w)\sim c(z)c(w)\sim0.
\end{equation}
The modes of these fields therefore satisfy the anticommutation relations
\begin{equation}
	\acomm{b_m}{c_n}=\delta_{m+n,0}{\wun}, \qquad \acomm{b_m}{b_n}=\acomm{c_m}{c_n}=0
\end{equation}
of the Lie superalgebra $\algbc$, where $m,n\in\ZZ+\frac{1}{2}$ in the \ns{} sector and $m,n\in\ZZ$ in the Ramond sector.
The energy momentum tensor is chosen to be
\begin{equation}
	T^{\supbc}(z)=\frac{1}{2} \brac[\big]{-\normord{b\,\pd c}(z)+\normord{\pd b\, c}(z)},
\end{equation}
corresponding to central charge $1$.  This gives both $b$ and $c$ conformal dimension $\frac{1}{2}$.  There is also a Heisenberg field $Q(z)=\normord{bc}(z)$ that gives $b$ and $c$ charges of $1$ and $-1$, respectively:
\begin{equation}
	Q(z)b(w)\sim \frac{b(w)}{z-w}, \qquad Q(z)c(w)\sim -\frac{c(w)}{z-w}.
\end{equation}

As with $N=2$ modules, we shall always assume that $\bcghost$-modules are $\ZZ_2$-graded.  Up to isomorphism, there are thus precisely four \hw{} $\algbc$-modules and all are simple: a \ns{} Verma module $\bcnsp$, a Ramond Verma module $\bcrp$, and their parity-reversals $\bcnsm = \parr \bcnsp$ and $\bcrm = \parr \bcrp$.  The vacuum module is $\bcnsp$.  The \hwv{} of the \ns{} modules has charge ($Q_0$-eigenvalue) $0$ and conformal dimension $0$.  We choose (arbitrarily) to regard $b_0$ as an annihilator and $c_0$ as a creator in the Ramond sector.  The \hwv{} of the Ramond modules thus has charge ($Q_0$-eigenvalue) $\frac{1}{2}$ and conformal dimension $\frac{1}{8}$.

As in any theory with fermions, it is appropriate to consider the character and supercharacter of a ($\ZZ_2$-graded) module $\bcmod{}$.  For fermionic ghosts, we define
\begin{equation}
	\fch{\bcmod{}}{x;q} = \traceover{\bcmod{}} x^{Q_0} q^{L^{\supbc}_0-1/24}, \qquad
	\fsch{\bcmod{}}{x;q} = \traceover{\bcmod{}} (-1)^F x^{Q_0} q^{L^{\supbc}_0-1/24},
\end{equation}
where $F \in \End(\bcmod{})$ acts as $0$ on the bosonic subspace and as \(1\) on the fermionic subspace.  The ghost characters and supercharacters are then easily verified to be given by
\begin{subequations}
	\begin{align}
		\fch{\bcnsp}{x;q} = \fch{\bcnsm}{x;q} &= \frac{\fjth{3}{x;q}}{\eta(q)}, &&&
		\fch{\bcrp}{x;q} = \fch{\bcrm}{x;q} &= \frac{\fjth{2}{x;q}}{\eta(q)},\label{chbc} \\
		\fsch{\bcnsp}{x;q} = -\fsch{\bcnsm}{x;q} &= \frac{\fjth{4}{x;q}}{\eta(q)}, &&&
		\fsch{\bcrp}{x;q} = -\fsch{\bcrm}{x;q} &= \frac{\ii \fjth{1}{x;q}}{\eta(q)}, \label{schbc}
	\end{align}
\end{subequations}
where $\jth{i}$ denotes the Jacobi theta functions, our conventions for which follow \cite[App.~B]{RidSL208}.

The fusion rules for the fermionic ghost modules can be deduced from those of the Heisenberg Fock spaces \eqref{eq:frH} by recalling that the former is an infinite-order simple current extension of the latter (this is the celebrated boson-fermion correspondence).  Indeed, restricting the ghost \svoa{} $\bcghost = \bcnsp$ to the Heisenberg subalgebra generated by $Q$ results in the branching rule
\begin{equation}
	\res{\bcnsp} \cong \bigoplus_{p \in \ZZ} \fock{p}.
\end{equation}
Taking into account the fact that the vectors in $\bcnsp$ with odd $Q_0$-charge are fermionic, we deduce that the Fock spaces induce to $\bcghost$-modules as follows:
\begin{equation}
	\ind{\fock{2n + i/2}} \cong \bcmod{i}, \quad n \in \ZZ.
\end{equation}
Here, we note that \(\fock{2n + i/2}\) is considered to be bosonic.
Using \cite[Eq.~(3.3)]{RidVer14}, which has been rigorously proven in \cite[Thm.~3.68]{CreTen17}, the fusion rules are now easily shown to be given by
\begin{equation}\label{eq:bcfus}
	\bcmod{i} \fuse \bcmod{j} \cong \ind{\fock{2n+i/2}} \fuse \ind{\fock{2m+j/2}} \cong \ind{\fock{2(m+n)+(i+j)/2}} \cong \bcmod{i+j},
\end{equation}
where the addition in the index of the final $\bcghost$-module is understood to be taken mod $4$.  Alternatively, these fusion rules can also be easily deduced from the fermionic Verlinde formula of \cite{EhoFus94,CanFusII15}.

Finally, the conjugation automorphism $\conjbcsymb$ and the spectral flow isomorphisms $\sfbcsymb^{\ell}$, $\ell \in \frac{1}{2} \ZZ$, of the ghost Lie superalgebra $\algbc$ have the following form (as usual, $\wun$ is left invariant by these automorphisms):
\begin{equation} \label{eq:bcauts}
	\begin{aligned}
		\conjbc{b_n} &= c_n, & \conjbc{c_n} &= b_n, & \conjbc{Q_n} &= -Q_n, & \conjbc{L^{\supbc}_n} &= L^{\supbc}_n, \\
		\sfbc{\ell}{b_n} &= b_{n-\ell}, & \sfbc{\ell}{c_n} &= c_{n+\ell}, & \sfbc{\ell}{Q_n} &= Q_n - \ell \delta_{n,0} \wun, & \sfbc{\ell}{L^{\supbc}_n} &= L^{\supbc}_n - \ell Q_n + \tfrac{1}{2} \ell^2 \delta_{n,0} \wun.
	\end{aligned}
\end{equation}
It is now easily verified that twisting the modules introduced above by these automorphisms leads to
\begin{equation}
	\conjbc{\bcmod{i}} \cong \bcmod{-i}, \qquad \sfbc{\ell}{\bcmod{i}} \cong \bcmod{i + 2 \ell},
\end{equation}
where we again understand that the ghost module indices are taken mod $4$.  Note that $\sfbcsymb^2$ is a (non-trivial) automorphism of each $\bcmod{i}$ while, up to isomorphism, $\sfbcsymb$ may be identified with the parity reversal functor $\parr$. Thus, as algebra isomorphisms, \(\conjbcsymb\) and \(\sfbcsymb^{1/2}\) generate the infinite dihedral group, while as twisting functors on isomorphism classes of modules they generate the symmetries of the square.

\subsection{The $\SLA{sl}{2}$ minimal models} \label{subsec:wzw}

The \voas{} associated with the affine Lie algebra $\algsl$ are generated by three bosonic fields, $e(z)$, $f(z)$ and $h(z)$, that satisfy the following \opes{}:
\begin{equation}
	\begin{gathered}
		h(z)e(w)\sim\frac{2e(w)}{z-w},\qquad h(z)h(w)\sim\frac{2k \wun}{(z-w)^2}, \qquad
		h(z)f(w)\sim\frac{-2f(w)}{z-w}, \\ \qquad e(z)f(w)\sim\frac{k \wun}{(z-w)^2}+\frac{h(w)}{z-w},\qquad
		e(z)e(w)\sim f(z)f(w)\sim 0.
	\end{gathered}
\end{equation}
Here, $k \in \CC \setminus \set{-2}$ is the \emph{level} of the \voa{}. The non-vanishing commutation relations between the modes of the generating fields are thus
\begin{equation}
	\comm{h_m}{e_n}=2e_{m+n},\quad\ \ \comm{h_m}{h_n}=2m\delta_{m+n,0}k \wun,\quad\ \ \comm{e_m}{f_n}=h_{m+n}+m\delta_{m+n,0}k \wun,\quad\ \ \comm{h_m}{f_n}=-2f_{m+n}.
\end{equation}

The energy-momentum tensor is given by the Sugawara construction:
\begin{equation}
	T^{\supsl}(z)=\frac{1}{2t}\left[\frac 1 2 \normord{hh}(z)+\normord{ef}(z)+\normord{fe}(z)\right].
\end{equation}
Here, $t = k+2 \in \CC \setminus \set{0}$ is the same parameter that appears in our conventions for the free boson (\cref{subsec:fb}).  The generating fields $e$, $f$ and $h$ are conformal primaries of conformal dimension $1$ and the central charge is
\begin{equation}
	\cc=3-\frac{6}{t}.
\end{equation}
The universal affine \voa{} associated to $\algsl$ is not simple if and only if there exist coprime $u\in\ZZ_{\ge2}$ and $v\in\ZZ_{\ge1}$ with $t = \frac{u}{v}$.  Its simple quotient will be referred to as an $\SLA{sl}{2}$ minimal model and we shall denote it by $\slminmod{u}{v}$.

The minimal models with $v=1$, hence non-negative integer levels $k$, are the \WZW{} models on the Lie group $\SLG{SU}{2}$.  They are also the \emph{unitary} minimal models of $\SLA{sl}{2}$.  Their irreducible modules are the integrable \hwms{} $\slirr{r}$, $r=1,\ldots,u-1$, whose characters are given by
\begin{equation}
	\fch{\slirr{r}}{w;q}=\traceover{\slirr{r}} w^{h_0}q^{L^{\supsl}_0-\cc/24}=\frac{q^{\slcdimuni{r}-\cc/24+1/8}}{\ii \vartheta_1(w^2;q)}\sum_{j\in\ZZ}\left(w^{2uj+r}-w^{-2uj-r}\right) q^{j(uj+r)}, \label{slcharuni}
\end{equation}
where $\slcdimuni{r} = \frac{1}{4u} (r^2-1)$ and we recall that we are using the conventions of \cite[App.~B]{RidSL208} for Jacobi theta functions.  Finally, the fusion rules are given by
\begin{equation}\label{eq:sl2fusuni}
\slirr{r}\fuse\slirr{r'}=\bigoplus_{r''=1}^{u-1}{\vircoe{u}{r''}{r,r'}}\slirr{r''},
\end{equation}
where the fusion coefficients are given by
\begin{equation} \label{eq:slfuscoeffs}
	\vircoe{u}{r''}{r,r'} =
	\begin{cases*}
	  1 & if $\abs{r-r'} +1 \leq r'' \le \min\set{r+r'-1, 2u-r-r'-1}$ and $r+r'+r''$ is odd, \\
		0 & otherwise.
	\end{cases*}
\end{equation}
The vacuum module is, of course, $\slirr{1}$.

When $v\ge2$, the minimal model $\slminmod{u}{v}$ is non-unitary, with fractional level $k = -2 + \frac{u}{v} \notin \ZZ$.  The irreducible positive-energy $\slminmod{u}{v}$-modules were classified in \cite{AdaVer95}, see also \cite{RidRel15}.  To facilitate the result, we introduce the following parametrisations for $r,s \in \ZZ$:
\begin{equation}
 \lambda_{r,s}=r-1-ts, \qquad \slcdim{r}{s}=\frac{(r-ts)^2-1}{4t}. \label{sl2weights}
\end{equation}
These can be easily checked to satisfy the identities
\begin{equation}
	\lambda_{u-r,v-s}=-\lambda_{r,s}-2, \qquad\slcdim{u-r}{v-s}=\slcdim{r}{s}.
\end{equation}
The list of (isomorphism classes of) irreducible positive-energy $\slminmod{u}{v}$-modules is then as follows:
\begin{itemize}
	\item The irreducible 	\hwms{} $\slirr{r,0}$, where $1\le r\le u-1$, whose \hwvs{} have $h_0$-charges $\lambda_{r,0}$ and conformal dimensions $\slcdim{r}{0}$.  Note that $\lambda_{r,0} = r-1 \in \ZZ_{\ge 0}$, hence the space of ground states is finite-dimensional.  The vacuum module is $\slirr{1,0}$.
	\item The irreducible \hwms{} $\sldis{r,s}$, where $1\le r\le u-1$ and $1\le s\le v-1$, whose \hwvs{} have charges $\lambda_{r,s}$ and conformal dimensions $\slcdim{r}{s}$. As $\lambda_{r,s} \notin \ZZ$, the space of 	ground states forms an infinite-dimensional irreducible Verma module for the horizontal subalgebra $\SLA{sl}{2}$.
	\item The irreducible modules $\sldism{r,s}$, where $1\le r\le u-1$ and $1\le s\le v-1$, that are conjugate to the $\sldis{r,s}$ (see \eqref{eq:slautcoincidences} below).  These are not \hwms{}. Indeed, the ground states of $\sldism{r,s}$ form an infinite-dimensional irreducible \lw{} Verma module over $\SLA{sl}{2}$ of \lw{} $-\lambda_{r,s}$.  The conformal dimension of each ground state is $\slcdim{r}{s}$.
	\item The irreducible \emph{relaxed} \hwms{} $\slrel{\lambda,\slcdim{r}{s}}$, where $1\le r\le u-1$, $1\le s\le v-1$ and $\lambda \in \CC$ satisfy $\lambda\neq\lambda_{r,s}, \lambda_{u-r,v-s} \bmod{2}$.  The ground states of $\slrel{\lambda,\slcdim{r}{s}}$ form an irreducible $\SLA{sl}{2}$-module that is neither highest- nor \lw{}.  The charges of the ground states are equal to $\lambda \bmod{2}$ and their conformal dimension is $\slcdim{r}{s}$.  There are isomorphisms $\slrel{\lambda,\slcdim{r}{s}} \cong \slrel{\lambda+2,\slcdim{r}{s}}$, for all $\lambda\neq\lambda_{r,s}, \lambda_{u-r,v-s} \bmod{2}$.
\end{itemize}

In addition to the irreducible modules listed above, there are also \emph{reducible} relaxed \hw{} $\slminmod{u}{v}$-modules corresponding to $\lambda = \lambda_{r,s}, \lambda_{u-r,v-s} \bmod{2}$.  In particular, there exist reducible $\slminmod{u}{v}$-modules $\slrelpm{r,s}$, where $1\le r\le u-1$ and $1\le s\le v-1$, whose ground states have charge equal to $\lambda_{r,s} \bmod{2}$ and conformal dimension $\slcdim{r}{s}$.  Moreover, $\slrelpm{r,s}$ is relaxed with a submodule isomorphic to $\sldispm{r,s}$ and the quotient by this submodule being isomorphic to $\sldismp{u-r,v-s}$ \cite{KawRel18}.  This is succinctly summarised in the following non-split short exact sequence:
\begin{equation} \label{eq:sessl2}
	\dses{\sldispm{r,s}}{}{\slrelpm{r,s}}{}{\sldismp{u-r,v-s}}.
\end{equation}

The characters of these $\slminmod{u}{v}$-modules are given by
\begin{subequations}\label{sl2char}
\begin{align}
\fch{\slirr{r,0}}{w;q}&=\frac{q^{\slcdim{r}{0}-\cc/24+1/8}}{\ii\vartheta_1(w^2;q)}\sum_{j\in\ZZ}\left(w^{2uj+r}-w^{-2uj-r}\right) q^{vj(uj+r)},\label{charL}\\
\fch{\sldispm{r,s}}{w;q}&=\frac{w^{\pm(\lambda_{r,s}+1)} q^{\slcdim{r}{s}-\cc/24+1/8}}{\pm\ii\vartheta_1(w^{2};q)}\sum_{j\in\ZZ}\left[w^{\pm2uj}q^{j(uvj+vr-us)}-w^{\pm2(uj-r)}q^{(uj-r)(vj-s)}\right],\label{charD}\\
\fch{\slrel{\lambda,\slcdim{r}{s}}}{w;q}&=\frac{w^{\lambda} \fchvir{r,s}}{{\eta(q)}^2} \delta(w^2),\qquad
\fch{\slrelpm{r,s}}{w;q}=\frac{w^{\lambda_{r,s}} \fchvir{r,s}}{{\eta(q)}^2} \delta(w^2), \label{charE}
\end{align}
\end{subequations}
where \(\delta(z)=\sum_{n\in\ZZ}z^n\) is the algebraic delta function and
\begin{equation}
    \chvir{r,s}(q)=\frac{1}{\eta(q)}\sum_{n\in\ZZ}\sqbrac*{q^{(2uvn+vr-us)^2/4uv}-q^{(2uvn+vr+us)^2/4uv}}
\end{equation}
denotes the character of the irreducible \hw{} Virasoro module whose conformal dimension is
\begin{equation}
\Delta^{\textup{Vir}}_{r,s}=\frac{(vr-us)^2-(v-u)^2}{4uv}.
\end{equation}
One must be careful with \eqref{sl2char} to expand the reciprocal of the theta function $\jth{1}$ in the correct annulus of convergence (in $w$), see \cite{RidSL208,CreMod13}.  The characters of the $\slrel{\lambda,\slcdim{r}{s}}$, which must be treated as distributions in $w$, were originally conjectured in \cite{CreMod13} and were subsequently proved in \cite{AdaRea17} (generically) and \cite{KawRel18} (in complete generality).

The Grothendieck fusion rules for the non-unitary minimal models $\slminmod{u}{v}$, $v \ge 2$, were computed in \cite{CreMod13} assuming the conjecture that the standard Verlinde formula of \cite{CreLog13,RidVer14} gives the Grothendieck fusion coefficients.  Based on these results, the actual fusion rules have also been conjectured \cite{CreCos18} and those involving just the $\slirr{r,0}$ have recently been proven \cite{CreBra17}.  These are all listed in \cref{app:frsl2}.

The conjugation ($\conjslsymb$) and spectral flow ($\sfslsymb^{\ell}$, $\ell \in \ZZ$) automorphisms of $\algsl$ preserve $\wun$, as usual, and act on the other generators as follows:
\begin{equation} \label{eq:slauts}
\begin{aligned}
\conjsl{e_n}&=f_n, & \conjsl{h_n}&=-h_n, & \conjsl{f_n}&=e_n,&  \conjsl{L^{\supsl}_0}&=L^{\supsl}_0,\\
\sfsl{\ell}{e_n}&=e_{n-\ell}, & \sfsl{\ell}{h_n}&=h_n-\ell\delta_{n,0} k \wun,&  \sfsl{\ell}{f_n}&=f_{n+\ell}, & \sfsl{\ell}{L^{\supsl}_0}&=L^{\supsl}_0-\frac 1 2 \ell h_0+\frac 1 4 \ell^2 k \wun.
\end{aligned}
\end{equation}
When $v=1$, the corresponding conjugation and spectral flow functors act on the irreducible modules as
\begin{equation}
	\conjsl{\slirr{r}} \cong \slirr{r}, \qquad \sfsl{}{\slirr{r}} \cong \slirr{u-r}.
\end{equation}
In general, the conjugates of the irreducible positive-energy $\slminmod{u}{v}$-modules are likewise easily identified:
\begin{equation} \label{eq:slautcoincidences}
	\conjsl{\slirr{r,0}} \cong \slirr{r,0}, \qquad
	\conjsl{\sldispm{r,s}} \cong \sldismp{r,s}, \qquad
	\conjsl{\slrel{\lambda,\slcdim{r}{s}}} \cong \slrel{-\lambda,\slcdim{r}{s}}, \qquad
	\conjsl{\slrelpm{r,s}} \cong \slrelmp{r,s}.
\end{equation}
However, spectral flow does not preserve the property of being positive-energy in general.  There are a small number of identifications, namely
\begin{equation}\label{eq:affineauto}
	\sfsl{\pm 1}{\slirr{r,0}} \cong \sldispm{u-r,v-1} \qquad \text{and} \qquad
	\sfsl{-1}{\sldis{r,s}} \cong \sldism{u-r,v-1-s} \quad \text{($s \neq v-1$)}
\end{equation}
and their obvious consequences, but in general applying spectral flow to an irreducible positive-energy $\slminmod{u}{v}$-module almost always results in an irreducible $\slminmod{u}{v}$-module whose conformal dimensions are not bounded below (the resulting module is therefore not usually positive-energy).  Consequently, conjugation and spectral flow generate the infinite dihedral group as algebra automorphisms, independent of the parameters \(u\) and \(v\), and this continues to be true for the corresponding functors if \(v>1\). However, if \(v=1\), then the group of twist functors, acting on isomorphism classes, collapses down to just \(\ZZ_2\).

In the language of the standard module formalism \cite{CreLog13,RidVer14} that describes the modular properties of the $\slminmod{u}{v}$-characters, for $v\ge2$, the $\slrel{\lambda,\slcdim{r}{s}}$ and $\slrelpm{r,s}$, together with their images under spectral flow, form the \emph{standard} modules of the minimal model $\slminmod{u}{v}$.  The irreducible standard modules, that is the $\slrel{\lambda,\slcdim{r}{s}}$, are said to be \emph{typical}, while the remaining indecomposable $\slminmod{u}{v}$-modules are said to be \emph{atypical}.  The latter class therefore includes the $\slirr{r,0}$, $\sldispm{r,s}$ and $\slrelpm{r,s}$, as well as their spectral flow images.  We shall use this terminology freely below, adapting it also to the non-unitary $N=2$ minimal models.

\subsection{The embedding} \label{embed}

The embedding \eqref{eq:embedding} of $\fboson$ and $\minmod{u}{v}$ as subalgebras of $\slminmod{u}{v}\otimes\bcghost$ is given explicitly, at the level of the generating fields, by
\begin{subequations} \label{eq:theembedding}
	\begin{gather}
		a(z)=h(z)+2Q(z),\label{phiinhbc} \\
		T^{\supsc}(z) = T^{\supsl}(z)+T^{\supbc}(z)-T^{\supH}(z)
		=\frac{1}{2t}\brac[\big]{\normord{ef}(z)+\normord{fe}(z)} - \frac{1}{t} h(z) Q(z) + \frac{k}{2t} \normord{QQ}(z), \label{Tembed}\\
		J(z)=\frac{1}{t}h(z)-\frac k t Q(z), \qquad
		G^+(z)=\sqrt{\frac{2}{t}}e(z)c(z), \qquad
		G^-(z)=\sqrt{\frac{2}{t}}f(z)b(z).\label{JGembed}
	\end{gather}
\end{subequations}
Because $a$ has regular \opes{} with $T^{\supsc}$, $J$, $G^+$ and $G^-$, this is actually an embedding of $\fboson \otimes \minmod{u}{v}$ into $\slminmod{u}{v}\otimes\bcghost$.

Of course, the identifications \eqref{Tembed} and \eqref{JGembed} by themselves do not prove that we have such an embedding.  Rather, they define a non-zero homomorphism of \svoas{} from the tensor product of $\fboson$ with the universal $N=2$ algebra to $\slminmod{u}{v}\otimes\bcghost$.  We therefore have an embedding of $\VOA{H} \otimes \VOA{V}$ into $\slminmod{u}{v}\otimes\bcghost$, where $\VOA{V}$ is some (indecomposable) quotient of the universal $N=2$ algebra.  As the zero modes $h_0$ and $Q_0$ act diagonalisably on $\slminmod{u}{v}$ and $\bcghost$, respectively, $a_0 = h_0 + 2Q_0$ acts diagonalisably on their tensor product.  It follows that $\slminmod{u}{v}\otimes\bcghost$ decomposes as an $\fboson \otimes \VOA{V}$-module as follows:
\begin{equation}
	\res{\brac*{\slminmod{u}{v}\otimes\bcghost}} \cong \bigoplus_{p \in 2\ZZ} \fock{p} \otimes \Mod{C}_p.
\end{equation}
Here, the $\Mod{C}_p$ are $\VOA{V}$-modules and, as $\fock{0} = \fboson$, the discussion above forces $\Mod{C}_0 = \VOA{V}$.  However, $\slminmod{u}{v}\otimes\bcghost$ is simple as a \svoa{}, since both $\slminmod{u}{v}$ and $\bcghost$ are, hence $\Mod{V}$ is simple by a result of Kac and Radul \cite[Thm.~1.1]{KacRep96} (see \cite[Sec.~3.1]{CreSch16} for a detailed discussion that puts this result into the context of cosets).  In other words, $\Mod{V} = \minmod{u}{v}$ and we have proven the desired embedding \eqref{eq:embedding}.

This simple proof stands in contrast to many of the arguments found in the literature.  One of the first arguments to address the simplicity of the coset \eqref{cosetN=2} is found in \cite{EhoUni96}, where it is established using explicit character computations.  However, this relied upon the Verma module embedding diagrams of \cite{DorSin95,DorEmb98} which are not universally acknowledged.  A different proof appears in \cite{AdaRep99}, based on the coset-inspired categorical equivalences sketched in \cite{FeiEqu98} but only recently proven in \cite{SatEqu16}.  Another proof, based on invariant theory, has recently appeared in \cite{CreCos14}.

The coset \eqref{cosetN=2} has implications for the conjugation and spectral flow automorphisms of $\minmod{u}{v}$, $\fboson$, $\bcghost$ and $\slminmod{u}{v}$, given in \cref{eq:N=2auts,eq:fbauts,eq:bcauts,eq:slauts}, respectively.  In particular, the following relationships are easily verified:
\begin{equation} \label{eq:cosetauts}
	\conjslsymb \otimes \conjbcsymb = \conjfbsymb \otimes \conjscsymb, \qquad
	\sfslsymb^{\ell} \otimes \sfbcsymb^{m} = \sffbsymb^{\ell k + 2m} \otimes \sfscsymb^{\ell-m}, \quad \ell \in \ZZ,\ m \in \tfrac{1}{2} \ZZ.
\end{equation}
The first merely states that conjugation is conserved by the coset.  The second is, however, quite powerful as we shall see.

\section{Branching rules} \label{sec:branch}

\subsection{Generalities}

Recall that an $N=2$ minimal model $\minmod{u}{v}$ is parametrised by two positive coprime integers $u \neq 1$ and $v$, which also describe the $\SLA{sl}{2}$ minimal model $\slminmod{u}{v}$ in the coset construction \eqref{cosetN=2}.  The minimal model $\minmod{u}{v}$ is unitary and rational when $v=1$ and is non-unitary and logarithmic otherwise.  We shall first study the consequences of the automorphism twist relations \eqref{eq:cosetauts} for the general branching rules.  This will enable us to easily analyse the branching rules of the unitary models, detailing the arguments in this familiar case, before generalising to those of their more involved non-unitary cousins.

The weight supports, meaning the sets of \(h_0\)-eigenvalues, of all the indecomposable ($k \neq 0$) \(\slminmod{u}{v}\)-modules are cosets in \(\CC/2\ZZ\), so let \(\Mod{M}_{\lambda}\) be such an indecomposable module with weight support \(\lambda+\CC/2\ZZ\). The field identifications \eqref{eq:theembedding} then imply that the eigenvalues of the Heisenberg zero mode \(a_0\) on \(\Mod{M}_{\lambda}\otimes \bcmod{i}\), $i=0,\dots,3$, lie in \(\lambda+i+2\ZZ\).  This means that the branching rule has the form
\begin{equation}\label{eq:generalbranchingrule}
  \res{(\Mod{M}_{\lambda} \otimes \bcmod{i})} \cong\ \bigoplus_{\mathclap{p \in\lambda+i+2\ZZ}}\ \fock{p} \otimes \tensor*[^{[i]}]{\Mod{C}}{_p^{\Mod{M}}},
\end{equation}
for some $\minmod{u}{v}$-modules \(\tensor*[^{[i]}]{\Mod{C}}{_p^{\Mod{M}}}\).  If \(\Mod{M}_{\lambda}\) is irreducible, then these $\minmod{u}{v}$-modules will be as well by \cite[Thm.~3.8]{CreSch16}.  Noting that the weight supports of \(\conjsl{\Mod{M}_{\lambda}}\) and \(\sfsl{\ell}{\Mod{M}_{\lambda}}\) are \(-\lambda+2\ZZ\) and \(\lambda+\ell k+2\ZZ=\lambda+\ell t+2\ZZ\), respectively, we can now derive many identifications among the \(N=2\) modules appearing in the branching rules of \(\Mod{M}_{\lambda}\) and its twists.

For example, putting $\ell=m=1$ into the second relation of \eqref{eq:cosetauts} gives \(\sfslsymb\otimes \sfbcsymb=\sffbsymb^t\otimes \wun_{\supsc}\).  Applying this to the branching rule \eqref{eq:generalbranchingrule} results in
\begin{equation}
  \bigoplus_{\mathclap{p \in\lambda+i+2\ZZ}}\ \fock{p+t}\otimes \tensor*[^{[i]}]{\Mod{C}}{_p^{\Mod{M}}}
  \cong \res{(\sfsl{}{\Mod{M}_{\lambda}} \otimes \sfbc{}{\bcmod{i}})}
  \cong \res{(\sfsl{}{\Mod{M}_{\lambda}} \otimes \bcmod{i+2})}
  \cong\ \bigoplus_{\mathclap{p \in\lambda+t+i+2\ZZ}}\ \fock{p}\otimes \tensor*[^{[i+2]}]{\Mod{C}}{_p^{\sfaut(\Mod{M)}}},
\end{equation}
where the \(\tensor*[^{[i+2]}]{\Mod{C}}{_p^{\sfaut(\Mod{M})}}\) are the \(N=2\) modules appearing in the branching rules of \(\sfsl{}{\Mod{M}_{\lambda}}\otimes \bcmod{i+2}\).
Thus,
\begin{equation} \label{eq:seedbrid}
  \tensor*[^{[i]}]{\Mod{C}}{_p^{\sfaut(\Mod{M})}}\cong \tensor*[^{[i+2]}]{\Mod{C}}{_{p-t}^{\Mod{M}}}.
\end{equation}
Similar identifications follow from applying \(\wun_{\text{aff.}}\otimes \sfbcsymb^{-\ell}=\sffbsymb^{-2\ell}\otimes \sfscsymb^{\ell}\) and \(\conjslsymb \otimes \conjbcsymb = \conjfbsymb \otimes \conjscsymb\) to \eqref{eq:generalbranchingrule}, which we summarise as follows:
\begin{equation}\label{eq:branchingids}
  \tensor*[^{[i]}]{\Mod{C}}{_p^{\sfaut^{\ell}(\Mod{M})}}\cong \tensor*[^{[i+2\ell]}]{\Mod{C}}{_{p-\ell t}^{\Mod{M}}},\qquad
  \sfsc{\ell}{\tensor*[^{[i]}]{\Mod{C}}{_p^{\Mod{M}}}}\cong \tensor*[^{[i-2\ell]}]{\Mod{C}}{_{p-2\ell}^{\Mod{M}}},\qquad
  \tensor*[^{[i]}]{\Mod{C}}{_p^{\conjaut(\Mod{M})}}\cong\conjsc{\tensor*[^{[-i]}]{\Mod{C}}{_{-p}^{\Mod{M}}}}.
\end{equation}
Note also that the coset preserves parity, so \(\tensor*[^{[i+2]}]{\Mod{C}}{_p^{\Mod{M}}}\cong\parr \tensor*[^{[i]}]{\Mod{C}}{_p^{\Mod{M}}}\).
We note, in particular, that the branching rules involving the spectral flows of \(\Mod{M}_{\lambda}\) produce no $\minmod{u}{v}$-modules that have not already appeared in the branching rules involving \(\Mod{M}_{\lambda}\).

\subsection{Unitary branching rules} \label{subsec:uniminmod}

Recall that the $\SLA{sl}{2}$ minimal models $\slminmod{u}{1}$ have precisely $u-1$ inequivalent irreducibles $\slirr{r}$, $r=1,\dots,u-1$, whose weight supports are $r-1+2\ZZ$.  We therefore arrive at the branching rules
\begin{equation} \label{eq:ubr}
	\res{(\slirr{r} \otimes \bcmod{i})} \cong\ \bigoplus_{\mathclap{p \in i+r-1+2\ZZ}}\ \fock{p} \otimes \scirrrnu{i}{p}{r},
\end{equation}
where the $\scirrrnu{i}{p}{r}$ are irreducible $\minmod{u}{1}$-modules.  Due to the fact that \(\sfsl{}{\slirr{r}}\cong \slirr{u-r}\), the $\scirrrnu{i}{p}{r}$ are not all inequivalent.  Indeed, \eqref{eq:seedbrid} implies that
\begin{equation}\label{eq:uKacSymm}
	\scirrrnu{i}{p}{r}\cong \scirrrnu{i+2}{p+u}{u-r}\cong\scirrrnu{i}{p+2u}{r},
	\qquad i=0,\dots,3,\ r=1,\dots,u-1,\ p\in i+r-1 +2\ZZ.
\end{equation}
The isomorphisms $\scirrrnu{i}{p}{r} \cong \scirrrnu{i}{p+2u}{r}$ imply that the commutant of $\minmod{u}{1} \cong \scirrrnu{0}{0}{1}$ in $\slminmod{u}{1} \otimes \bcghost$ is not $\fboson \cong \fock{0}$, but is rather the lattice \voa{} (compactified free boson) $\mathbb{F}_{2u} \cong \bigoplus_{p \in 2u\ZZ} \fock{p}$.  Accordingly, the branching rules \eqref{eq:ubr} may be rewritten as direct sums over tensor products of $\mathbb{F}_{2u}$- and $\minmod{u}{1}$-modules.  These facts can be useful in many ways, in particular as $\mathbb{F}_{2u}$ is rational, but will not be required in what follows.

In any case, the total number of inequivalent irreducible \hw{} $\minmod{u}{1}$-modules that we have obtained is bounded above by $2u(u-1)$.  To better understand the coset modules $\scirrrnu{i}{p}{r}$ and show that there are no further isomorphisms between them, we need to identify them with the $N=2$ modules $\NSRIrr{\pm}{j}{\Delta}$ introduced in \cref{sec:N=2Reps}.  This dictionary between the two notations is easily constructed using the method of extremal states.

The \emph{extremal states} of a module are defined to be those states which, for a given fixed charge, have the minimal possible conformal dimension.  In the case at hand, the extremal states are the minimal conformal dimension states of $\slirr{r} \otimes \bcmod{i}$ in each subspace of constant $a_0$-charge, where we recall from \eqref{phiinhbc} that $a_0 = h_0 + 2 Q_0$.  The minimality condition ensures that such a state is necessarily annihilated by the positive modes of $\fboson$ and $\minmod{u}{1}$.  As both $\fock{p}$ and $\scirrrnu{i}{p}{r}$ are irreducible, they may be identified by computing the $a_0$-, $J_0$- and $L^{\supsc}_0$-eigenvalues of the \hw{} extremal states.

To illustrate, we consider $\slirr{r} \otimes \bcmod{0}$.  Its extremal states may be readily found as a subset of the states obtained by tensoring an extremal state of $\slirr{r}$ with one of $\bcmod{0}$.  Let $\ket{r}$ and $\ket{\NS^+}$ denote the \hwvs{} of $\slirr{r}$ and $\bcmod{0}$, respectively, recalling that the $h_0$-charge of $\ket{r}$ is $r-1$ and the $Q_0$-charge of $\ket{\NS^+}$ is $0$.  The extremal states of $\slirr{r}$ and $\bcmod{0}$ include
\begin{equation}
	\begin{aligned}
		f_0^m &\ket{r} & &\text{($m=0,1,\dots,r-1$),} \\
		e_{-1}^n &\ket{r} & &\text{($n=0,1,\dots,u-r-1$),}
	\end{aligned}
	\qquad b_{-1/2}\ket{\NS^+},\quad \ket{\NS^+}, \quad c_{-1/2} \ket{\NS^+}
\end{equation}
(there are many others, but these will suffice for our analysis).  In $\slirr{r} \otimes \bcmod{0}$, minimising conformal dimensions now easily verifies that the extremal state of $a_0$-charge $r-1-2m$, $m=0,1,\dots,r-1$, has the form $f_0^m \ket{r} \otimes \ket{\NS^+}$ and that of $a_0$-charge $r+1+2n$, $n=0,1,\dots,u-r-1$, has the form $e_{-1}^n \ket{r} \otimes b_{-1/2} \ket{\NS^+}$.  The former are therefore bosonic with conformal dimension $\slcdimuni{r}$ while the latter are fermionic with conformal dimension $\slcdimuni{r} + n + \frac{1}{2}$.

Identifying these extremal states as \hwvs{} of $\fock{p} \otimes \scirrrnu{0}{p}{r}$, with $p=r-1-2m$ or $p=r+1+2n$, we use
\begin{equation}
	J_0 = \frac{h_0 - k Q_0}{u} \quad \text{and} \quad L^{\supsc}_0 = L^{\supsl}_0 + L^{\supbc}_0 - L^{\supH}_0
\end{equation}
to identify the irreducible $\minmod{u}{1}$-modules that they generate.  In this way, we find that the dictionary between the coset and $N=2$ notations for these modules is given by
\begin{subequations} \label{eq:unitarydictionary}
	\begin{equation} \label{eq:nsunitarydictionary}
		\scirrrnu{0}{p}{r} \cong \NSIrr{\bullet}{j}{\Delta}, \quad \text{where}\ \left\{
		\begin{aligned}
			\bullet &= +, & j &= \frac{p}{u}, & \Delta&=\sccdimuni{p}{r}, & \text{(} p &= -r+1,\dots,r-1 \text{),} \\
			\bullet &= -, & j &= \frac{p}{u}-1, & \Delta&=\sccdimuni{p}{r} + \frac{p-r}{2}, & \text{(} p &= r+1,\dots,2u-r-1 \text{),}
		\end{aligned}
		\right.
	\end{equation}
where $\sccdimuni{p}{r}=\slcdimuni{r,0}-\frac{p^2}{4u}$. This identification must be supplemented by $\scirrrnu{0}{p}{r} \cong \scirrrnu{0}{p \pm 2u}{r}$, if $p \in r-1+2\ZZ$ does not fall in the range $-r+1,\dots, 2u-r-1$.  The dictionary for Ramond modules is similarly found to be
	\begin{equation} \label{eq:runitarydictionary}
		\scirrrnu{1}{p}{r} \cong \RIrr{\bullet}{j}{\Delta}, \quad \text{where}\ \left\{
		\begin{aligned}
			\bullet &= -, & j &= \frac{p}{u}+\frac{1}{2}, & \Delta&=\sccdimuni{p}{r} +\frac{1}{8}, & \text{(} p &= -r,\dots,r-2 \text{),} \\
			\bullet &= +, & j &= \frac{p}{u}-\frac{1}{2}, & \Delta&=\sccdimuni{p}{r} +\frac{1}{8} + \frac{p-r}{2}, & \text{(} p &= r,\dots,2u-r-2 \text{).}
		\end{aligned}
		\right.
	\end{equation}
\end{subequations}
Again, if $p \in r+2\ZZ$ does not fall in the range $-r,\dots 2u-r-2$, then we have $\scirrrnu{1}{p}{r} \cong \scirrrnu{1}{p \pm 2u}{r}$. The dictionaries for $i=2$ and $3$ are obtained from those for $i=0$ and $1$, respectively, by reversing parities.

We remark that if $p$ and $r$ satisfy $-r \le p \le r-1$, then $p+u$ and $u-r$ satisfy $u-r \le u+p \le 2u-(u-r)-1$.  In other words, the two branches of each dictionary are exchanged under the isomorphism \eqref{eq:uKacSymm}.  It follows that we may restrict to a single branch, say that for $-r \le p \le r-1$, remembering that the other just corresponds to its parity-reversal.  We therefore have a uniform parametrisation for the irreducible $\minmod{u}{1}$-modules obtained through the coset construction:
\begin{equation} \label{eq:unitaryparametrisation}
	\NSRIrr{\pm}{j}{\Delta}, \qquad
	\begin{aligned}
		r &= 1,2,\dots,u-1, \\
		p &= -r,-r+1,\dots r-1,
	\end{aligned}
	\qquad
	\begin{aligned}
		j=j_{p,r} &= \frac{p}{u} + \frac{1 + (-1)^{p+r}}{4}, \\
		\Delta=\sccdimuni{p}{r} &= \frac{r^2-p^2-1}{4u} + \frac{1 + (-1)^{p+r}}{16}.
	\end{aligned}
\end{equation}
The module is \ns{} for $p+r$ odd and Ramond for $p+r$ even.

It follows from the formula for $j_{p,r}$ that modules with different parameters $p$ in \eqref{eq:unitaryparametrisation} are not isomorphic.  Comparing $\sccdimuni{p}{r}$ and $\sccdimuni{p}{r'}$ now shows that the modules in \eqref{eq:unitaryparametrisation} are all distinct, hence that the coset construction produces precisely $2u(u-1)$ inequivalent irreducible $\minmod{u}{1}$-modules (including parity).  In fact, it is easy to show that there can be no more than $2u(u-1)$.  This relies on the result \cite[Thm.~4.3]{CreSch16} that given any irreducible $\minmod{u}{1}$-module $\Mod{C}$, one can find a Fock space $\fock{p}$ such that $\fock{p} \otimes \Mod{C}$ may be induced to an $\slminmod{u}{1} \otimes \bcghost$-module using the embedding \eqref{eq:embedding}.  The induced module will then decompose as a direct sum of irreducibles $\Mod{M} \otimes \bcmod{i}$, meaning that each $\Mod{M}$ is an irreducible $\slminmod{u}{1}$-module, and thus $\fock{p} \otimes \Mod{C}$ will appear in the branching rule of at least one of the $\Mod{M} \otimes \bcmod{i}$.  However, we have determined the branching rules for a complete set of irreducible $\slminmod{u}{1}$-modules, so $\Mod{C}$ must be one of the $2u(u-1)$ irreducible $\minmod{u}{1}$-modules identified above.

One can arrange the identifying data of the irreducibles $\scirrrnu{i}{p}{r}$ into a table reminiscent of the Kac table of the Virasoro minimal models.  We label the rows of this Kac table by $r=1,\dots,u-1$ and the columns by $p=-r,\dots,2u-r-1$, illustrating it for $\minmod{4}{1}$ in \cref{k2Kac} (top).  We note that the isomorphisms \eqref{eq:uKacSymm} allow us to reduce this table by half at the cost of ignoring parity information.  We also indicate this reduced table for $\minmod{4}{1}$ in \cref{k2Kac} (bottom).

\begin{table}
	\renewcommand{\arraystretch}{1.5} 
	\scalebox{0.85}{
		\begin{tabular}{C|C|*{10}{C}|}  
			\multicolumn{1}{C}{} & \multicolumn{1}{C}{} & \multicolumn{10}{C}{p} \\
			\cline{2-12}
			\multicolumn{1}{C|}{} & \pm;j;\Delta & -3 & -2 & -1 & 0 & 1 & 2 & 3 & 4 & 5 & 6 \\
			\cline{2-12}
			\multirow{3}{*}{$r$} & 1 &&& \RC -;\frac{1}{4};\frac{1}{16} & +;0;0 & \RC +;-\frac{1}{4};\frac{1}{16} & -;-\frac{1}{2};\frac{1}{4} & \RC +;\frac{1}{4};\frac{9}{16} & -;0;\frac{1}{2} & \RC +;\frac{3}{4};\frac{9}{16} & -;\frac{1}{2};\frac{1}{4} \\
			                   & 2 && \RC -;0;\frac{1}{16} & +;-\frac{1}{4};\frac{1}{8} & \RC -;\frac{1}{2};\frac{5}{16} & +;\frac{1}{4};\frac{1}{8} & \RC +;0;\frac{1}{16} & -;-\frac{1}{4};\frac{1}{8} & \RC +;\frac{1}{2};\frac{5}{16} & -;\frac{1}{4};\frac{1}{8} & \\
			                   & 3 & \RC -;-\frac{1}{4};\frac{1}{16} & +;-\frac{1}{2};\frac{1}{4} & \RC -;\frac{1}{4};\frac{9}{16} & +;0;\frac{1}{2} & \RC -;\frac{3}{4};\frac{9}{16} & +;\frac{1}{2};\frac{1}{4} & \RC +;\frac{1}{4};\frac{1}{16} & -;0;0 && \\
			\cline{2-12}
		\end{tabular}
	}
	\\[5mm]
	\begin{tabular}{C|C|*{6}{C}|}
		\multicolumn{1}{C}{} & \multicolumn{1}{C}{} & \multicolumn{6}{C}{p} \\
		\cline{2-8}
		\multicolumn{1}{C|}{} & j;\Delta & -3 & -2 & -1 & 0 & 1 & 2 \\
		\cline{2-8}
		\multirow{3}{*}{$r$} & 1 &&& \RC \frac{1}{4};\frac{1}{16} & 0;0 && \\
		                   & 2 && \RC 0;\frac{1}{16} & -\frac{1}{4};\frac{1}{8} & \RC \frac{1}{2};\frac{5}{16} & \frac{1}{4};\frac{1}{8} & \\
		                   & 3 & \RC -\frac{1}{4};\frac{1}{16} & -\frac{1}{2};\frac{1}{4} & \RC \frac{1}{4};\frac{9}{16} & 0;\frac{1}{2} & \RC \frac{3}{4};\frac{9}{16} & \frac{1}{2};\frac{1}{4} \\
		\cline{2-8}
	\end{tabular}
	\caption{The Kac table of $\minmod{4}{1}$ ($\cc=\frac{3}{2}$).  At top is the full table, where each irreducible module is labelled by its parity $\pm$, its charge $j$ and its conformal dimension $\Delta$.  The sector is indicated by shading Ramond cells.  At bottom is a simplified Kac table in which parity is ignored and the ``Kac symmetry'' \eqref{eq:uKacSymm} is used to remove half the modules.  The charges and conformal dimensions in this pyramidal table are computed using \eqref{eq:unitaryparametrisation}.}
	\label{k2Kac}
\end{table}

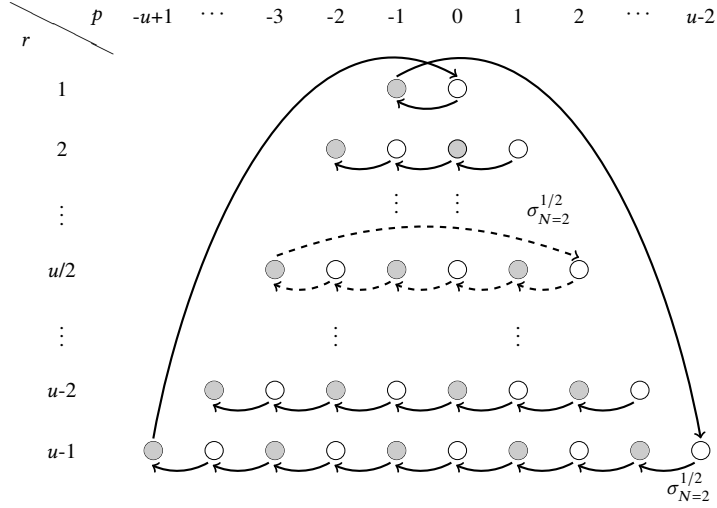
\begin{figure}
	\scalebox{0.8}{
		\begin{tikzpicture}
		\node at (-0.5,5) {\backslashbox[3mm]{$r$}{$p$}}; \node at (1,5.2) {-$u$+1}; \node at (2,5.2) {$\cdots$};
		\node at (3,5.2) {-3}; \node at (4,5.2) {-2}; \node at (5,5.2) {-1}; \node at (6,5.2) {0};
		\node at (7,5.2) {1}; \node at (8,5.2) {2}; \node at (9,5.2) {$\cdots$}; \node at (10,5.2) {$u$-2};
		\node at (-0.5,4) {1};
		\draw (6,4) circle (1.5mm);
		\draw (5,4) circle (1.5mm); \fill[gray!40!white] (5,4) circle (1.5mm);
		\node at (-0.5,3) {2};
		\draw (4,3) circle (1.5mm); \fill[gray!40!white] (4,3) circle (1.5mm);
		\draw (5,3) circle (1.5mm);\fill[gray!40!white] (6,3) circle (1.5mm);
		\draw (6,3) circle (1.5mm);
		\draw (7,3) circle (1.5mm);
		\node at (-.5,2){$\vdots$}; \node at (5,2.2){$\vdots$}; \node at (6,2.2){$\vdots$};
		\node at (-.5,1){$u$/2};
		\foreach \a in{3,4,...,8}
		\draw (\a,1) circle (1.5mm);
		\foreach \a in{3,5,7}
		\fill[gray!40!white] (\a,1) circle (1.5mm);
		\foreach \a in{3.9,4.9,...,7.9}
		\draw [dashed, ->, line width=1pt] (\a,0.8)  arc (-60:-115:1);
		\draw [dashed, ->, line width=1pt] (3.8,2,2) to [out=20,in=160] (8,1.2);
		\node at (-0.5,0){$\vdots$}; \node at (4,0){$\vdots$}; \node at (7,0){$\vdots$};
		\node at (-0.5,-1){$u$-2};
		\foreach \a in{2,3,...,9}
		\draw (\a,-1) circle (1.5mm);
		\foreach \a in{2,4,6,8}
		\fill[gray!40!white] (\a,-1) circle (1.5mm);
		\node at (-.5,-2){$u$-1};
		\foreach \a in{1,2,3,...,10}
		\draw (\a,-2) circle (1.5mm);
		\foreach \a in{1,3,5,7,9}
		\fill[gray!40!white] (\a,-2) circle (1.5mm);
		\foreach \a in{9.9, 8.9, ..., 1.9}
		\draw [->, line width=1pt] (\a, -2.2)  arc (-60:-115:1);
		\node at (9.8,-2.6) {$\sfscsymb^{1/2}$};
		\node at (7.5,2) {$\sfscsymb^{1/2}$};
		\draw [->, line width=1pt]  (1,-1.8) to [out=80,in=150] (6,4.2);
		\draw [->, line width=1pt] (6,3.8)  arc (-60:-120:1);
		\draw [->, line width=1pt]  (5,4.2) to [out=30,in=100] (10,-1.8);
		\foreach \a in{8.9, 7.9, ..., 2.9}
		\draw [->, line width=1pt] (\a, -1.2)  arc (-60:-115:1);
		\foreach \a in{4.9,5.9,6.9}
		\draw [->, line width=1pt] (\a, 2.8)  arc (-60:-115:1);
		\end{tikzpicture}
	}
	\caption{Spectral flow acting on the Kac table of the reduced minimal model $\minmod{u}{1}$. The white/grey circles represent \ns{}/Ramond modules, respectively. The $n$th and the $n$th-last rows together form a single orbit.  When $u$ is even, there is a middle row in the Kac table which forms a closed orbit on its own.}\label{n2sf}
\end{figure}

\begin{figure}
	\scalebox{0.8}{
		\begin{tikzpicture}
		\node at (-0.5,5) {\backslashbox[3mm]{$r$}{$p$}}; \node at (1,5.2) {-$u$+1}; \node at (2,5.2) {$\cdots$};
		\node at (3,5.2) {-3}; \node at (4,5.2) {-2}; \node at (5,5.2) {-1}; \node at (6,5.2) {0};
		\node at (7,5.2) {1}; \node at (8,5.2) {2}; \node at (9,5.2) {$\cdots$}; \node at (10,5.2) {$u$-2};
		\node at (-.5,4) {1}; \draw (6,4) circle (1.5mm); \draw (5,4) circle (1.5mm); \fill[gray!40!white] (5,4) circle (1.5mm);
		\node at (-.5,3) {2}; \draw (4,3) circle (1.5mm); \fill[gray!40!white] (4,3) circle (1.5mm);\draw (5,3) circle (1.5mm);\fill[gray!40!white] (6,3) circle (1.5mm);\draw (6,3) circle (1.5mm);\draw (7,3) circle (1.5mm);
		\node at (-.5,2){$\vdots$}; \node at (3,2.2){$\iddots$};\node at (5,2.4){$\vdots$}; \node at (7,2.4){$\vdots$};
		\node at (-.5,1){$u$-2};
		\foreach \a in{2,3,...,9}
		\draw (\a,1) circle (1.5mm);
		\foreach \a in{2,4,6,8}
		\fill[gray!40!white] (\a,1) circle (1.5mm);
		\node at (-0.5,0){$u$-1};
		\foreach \a in{1,2,3,...,10}
		\draw (\a,0) circle (1.5mm);
		\foreach \a in{1,3,...,9}
		\fill[gray!40!white] (\a,0) circle (1.5mm);
		\draw[dashed] (6,4.5)--(10.7,-0.3)--(1.3,-0.3)--(6,4.5);
		\draw [<->, line width=1pt]  (2,1.2) to [out=70,in=-160] (3.8,3.2);
		\draw [<->, line width=1pt]  (3,-0.3) to [out=-30,in=-150] (9,-0.3);
		\draw [<->, line width=1pt] (6.15,1.2)  arc (-45:240:0.3);
		\node at (6,-0.8) {$\conjscsymb$};\node at (5.2,1.5) {$\conjscsymb$};\node at (1.8,2.2) {$\conjscsymb$};
		\end{tikzpicture}
	}
	\caption{The action of conjugation on the reduced Kac table of the minimal model $\minmod{u}{1}$. In the dashed triangle, conjugation is effected by reflection about the central column ($p=0$). The modules outside this triangle form a strip on which conjugation is effected by reflection about the strip's middle point.}\label{n2conj}
\end{figure}
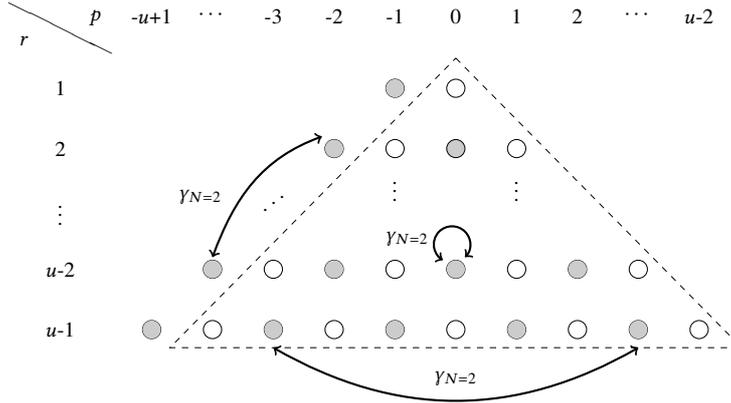

In \cref{sec:algprelim} we described conjugation and spectral flow as isomorphisms of the \(N=2\) super conformal algebras, so let us now analyse their action on the irreducible $\minmod{u}{1}$-modules. Applying the conjugation identity of \eqref{eq:cosetauts} to the branching rules \eqref{eq:ubr} immediately implies that \(\conjsc{\scirrrnu{i}{p}{r}}\cong \scirrrnu{-i}{-p}{r}\).  Similarly, setting \(\ell=0\) and \(m=-1/2\) in the spectral flow identity implies that
\begin{equation}\label{eq:sfrelunit}
\sfsc{1/2}{\scirrrnu{i}{p}{r}}\cong \scirrrnu{i-1}{p-1}{r}.
\end{equation}
It is then not hard to verify, by inspection, that spectral flow and conjugation partition the $2u(u-1)$ simple modules into orbits under the action of these automorphisms. The number of orbits and the orbit lengths depend on the parameter $u$ of the minimal model. This is summarised in the following list:
\begin{itemize}
	\item There are $u$ orbits when $u \in 4\ZZ+2$, two for each $r=1,\dots,\frac{u}{2}$ (one being the parity reversal of the other).  For $r < \frac{u}{2}$, the orbit length is $2u$, but for $r = \frac{u}{2}$, the orbit length is only $u$.  Representatives for these orbits are the $\scirrrnu{r\pm1}{0}{r}$, for $r<\frac{u}{2}$, and $\scirrrnu{u/2\pm1}{0}{u/2}$, for $r=\frac{u}{2}$.
	\item There are $u-1$ orbits when $u \in 4\ZZ$, with two for each $r=1,\dots,\frac{u}{2}-1$ (one being the parity reversal of the other) but only one for $r=\frac{u}{2}$ (closed under parity reversal).  All orbits have length $2u$.  Representatives for these orbits are the $\scirrrnu{r\pm1}{0}{r}$, for $r<\frac{u}{2}$, and $\scirrrnu{u/2-1}{0}{u/2}$, for $r=\frac{u}{2}$.
	\item There are $\frac{u-1}{2}$ orbits when $u \in 2\ZZ+1$, one for each $r=1,\dots,\frac{u-1}{2}$, all of length $4u$ and all closed under parity reversal.  Representatives for these orbits are the $\scirrrnu{r-1}{0}{r}$.
\end{itemize}
This is easily deduced from the fact that $\sfscsymb^{1/2}$ will only change the parity of an irreducible $\minmod{u}{1}$-module if it is Ramond with $\Delta \neq \frac{\cc}{24}$.  We illustrate these orbits pictorially for the \emph{reduced} Kac tables in \cref{n2sf,n2conj}.

\subsection{Non-unitary branching rules}

In this section, we shall generalise the extremal state method described in the study of $\minmod{u}{1}$ to the non-unitary $N=2$ minimal models $\minmod{u}{v}$, where $u,v\in\ZZ_{\ge2}$ and $\gcd\set{u,v}=1$. Following the coset \eqref{cosetN=2}, we construct different types of coset modules by decomposing the different $\slminmod{u}{v}$-modules, introduced in \cref{subsec:wzw}, tensored with $\bcghost$-modules.  We start with the irreducible $\slminmod{u}{v}$-modules, for which the resulting branching rules have the form
\begin{subequations} \label{eq:nubr}
\begin{gather}
	\res{(\slirr{r,0} \otimes \bcmod{i})} \cong\ \bigoplus_{\mathclap{p \in i+\lambda_{r,0}+2\ZZ}}\ \fock{p} \otimes {\scirrrn{i}{p}{r,0}}, \qquad
	 \res{(\sldis{r,s} \otimes \bcmod{i})} \cong\ \bigoplus_{\mathclap{p \in i+\lambda_{r,s}+2\ZZ}}\ \fock{p} \otimes \scdisrn{i}{p}{r,s}, \label{eq:Ddecomp}\\
	\res{(\slrel{\lambda,\slcdim{r}{s}} \otimes \bcmod{i})} \cong\ \bigoplus_{\mathclap{p \in i+\lambda+2\ZZ}}\ \fock{p} \otimes \screlrn{i}{p}{r,s},
\end{gather}
\end{subequations}
where $1\le r\le u-1$, $1 \le s\le v-1$, $i=0,\dots,3$ and $\lambda\neq\lambda_{r,s}, \lambda_{u-r,v-s} \bmod{2}$.  The $\scirrrn{i}{p}{r,0}$, $\scdisrn{i}{p}{r,s}$ and $\screlrn{i}{p}{r,s}$ are then irreducible $\minmod{u}{v}$-modules, by \cite[Thm.~3.8]{CreSch16}.

Note that it is not necessary to consider the branching rules involving the \(\sldism{r,s}\) because they are spectral flow images of the $\sldis{r,s}$, hence will not produce new \(\minmod{u}{v}\)-modules.  This also applies to the $\slirr{r,0} = \sfsl{-1}{\sldis{u-r,v-1}}$, so we shall implicitly exclude the branching rules corresponding to the $\sldis{r,v-1}$ in what follows.  In contrast to the unitary case, an $\slminmod{u}{v}$-module is never isomorphic to any of its non-trivial spectral flow twists \cite{CreMod13}, so the periodicity condition \eqref{eq:uKacSymm} is no longer valid in the non-unitary case.

By identifying the extremal states of $\slminmod{u}{v}\otimes\bcghost$-modules as \hwvs{} for the action of $\fboson\otimes\minmod{u}{v}$, we identify the infinitely many inequivalent irreducible \(\minmod{u}{v}\)-modules in these branching rules as modules of the \ns{} or Ramond $N=2$ Lie superalgebra.  The dictionary for identifying $\Mod{L}$-type \(\minmod{u}{v}\)-modules is
\begin{subequations}\label{eq:Lid}
	\begin{align}
		{\scirrrn{0}{p}{r,0}} &\cong \NSIrr{\bullet}{j}{\Delta}, & p &\in \lambda_{r,0}+2\ZZ, & &
		\begin{cases}
			\bullet = -,\  j = \frac{p}{t}+1,\ \Delta = \sccdim{p}{r}{0}-\frac{p+r}{2}, & p\le -r-1 \\
			\bullet = +,\  j = \frac{p}{t},\phantom{+1}\ \,\ \Delta = \sccdim{p}{r}{0}, & 1-r\le p\le r-1, \\
			\bullet = -,\  j = \frac{p}{t}-1,\ \Delta = \sccdim{p}{r}{0}+\frac{p-r}{2}, & p\ge r+1,
		\end{cases}
		\label{eq:nsNonunitarydictionaryL} \\
		{\scirrrn{1}{p}{r,0}} &\cong \RIrr{\bullet}{j}{\Delta}, & p &\in \lambda_{r,0}+1+2\ZZ, & &
		\begin{cases}
			\bullet = +,\  j = \frac{p}{t}+\frac 3 2 ,\  \Delta = \sccdim{p}{r}{0}+\frac{1}{8}-\frac{p+r}{2}, & p\le -r-2,\\
			\bullet = -,\  j = \frac{p}{t}+\frac 1 2,\  \Delta = \sccdim{p}{r}{0}+\frac{1}{8}, & -r\le p\le r-2, \\
			\bullet = +,\  j = \frac{p}{t}-\frac 1 2,\  \Delta = \sccdim{p}{r}{0}+\frac{1}{8}+\frac{p-r}{2}, & p\ge r,
		\end{cases}
		\label{eq:rNonunitarydictionaryL}
	\end{align}
\end{subequations}
where
\begin{equation} \label{eq:n2cdimsymm}
	\sccdim{p}{r}{s}=\slcdim{r}{s}-\frac{p^2}{4t}.
\end{equation}
Note that when comparing the dictionary for the Ramond with that of the \ns{} modules, the $J_0$-charges and conformal dimensions are shifted by $\frac{1}{2}$ and $\frac 1 8$, respectively, while parities are reversed.  The dictionaries for the $\Mod{D}$- and $\Mod{E}$-type irreducibles are as follows:
\begin{subequations}\label{eq:Did}
  \begin{align}
    \scdisrn{0}{p}{r,s} &\cong \NSIrr{\bullet}{j}{\Delta}, & p &\in \lambda_{r,s}+2\ZZ, & &
		\begin{cases}
			\bullet = +,\  j = \frac{p}{t},\phantom{+1}\ \ \, \Delta = \sccdim{p}{r}{s}, & p\le \lambda_{r,s}, \\
			\bullet = -,\  j = \frac{p}{t}-1,\  \Delta =\sccdim{p}{r}{s}+\frac{p-\lambda_{r,s}-1}{2}, & p\ge \lambda_{r,s}+2 ,
		\end{cases}
		\label{eq:nsNonunitarydictionaryDp} \\
    \scdisrn{1}{p}{r,s} &\cong \RIrr{\bullet}{j}{\Delta}, & p &\in \lambda_{r,s}+1+2\ZZ, & &
    \begin{cases}
      \bullet = -,\  j = \frac{p}{t}+\frac 1 2,\  \Delta = \sccdim{p}{r}{s}+\frac{1}{8}, & p\le \lambda_{r,s}-1,\\
      \bullet = +,\  j = \frac{p}{t}-\frac 1 2,\  \Delta =\sccdim{p}{r}{s}+\frac{1}{8}+\frac{p-\lambda_{r,s}-1}{2}, & p\ge\lambda_{r,s}+1,\\
    \end{cases}
    \label{eq:rNonunitarydictionaryDp}
  \end{align}
\end{subequations}
\begin{equation}
	\screlrn{0}{p}{r,s} \cong \NSIrr{+}{p/t}{\sccdim{p}{r}{s}},	\quad p\in \lambda+2\ZZ, \qquad
	\screlrn{1}{p}{r,s} \cong \RIrr{-}{p/t+1/2}{\sccdim{p}{r}{s}+1/8}, \quad p\in \lambda+1+2\ZZ.
	\label{eq:Eid}
\end{equation}
As usual, the dictionaries for $i=2$ and $3$ are obtained from those with $i=0$ and $1$, respectively, by reversing parities.

In addition, we can similarly deduce the branching rules of the reducible indecomposable $\slminmod{u}{v}$-modules $\slrelpm{r,s}$. Since the weight support of $\slrelpm{r,s}$ is \(\lambda_{r,s}+2\ZZ\), the branching rules have the form
\begin{equation} \label{br:atypst}
	\res{(\slrelpm{r,s} \otimes \bcmod{i})} \cong\ \bigoplus_{\mathclap{p \in \lambda_{r,s}+i+2\ZZ}}\ \fock{p} \otimes \scarelrn{i}{p}{r,s}.
\end{equation}
Since restriction is exact, combining the exact sequences \eqref{eq:sessl2} with the identifications \eqref{eq:branchingids} gives the following exact sequences for the $\scarelrn{i}{p}{r,s}$:
\begin{equation} \label{eq:sesN2}
	\begin{gathered}
	  \dses{\scdisrn{i}{p}{r,s}}{}{\scarelrnp{i}{p}{r,s}}{}{\scdisrn{i+2}{p+t}{r,s-1}},\\
	  \dses{\scdisrn{i+2}{p+t}{u-r,v-s-1}}{}{\scarelrnm{i}{p}{r,s}}{}{\scdisrn{i}{p}{u-r,v-s}}.
	\end{gathered}
\end{equation}
In these sequences, any occurrence of \(\scdisrn{i}{p}{r,0}\) should be replaced by \(\scirrrn{i}{p}{r,0}\). In analogy to the nomenclature introduced for \(\slminmod{u}{v}\), the $\minmod{u}{v}$-modules \(\screlrn{i}{p}{r,s}\) and \(\scarelrn{i}{p}{r,s}\) will be referred to as \emph{standard modules}. Further, the modules \(\screlrn{i}{p}{r,s}\) will be referred to as being \emph{typical} while all other indecomposable modules will be referred to as \emph{atypical}.

Recall from \eqref{eq:n2cdimsymm} that the conformal dimension and charge $j=p/t$ of the \hwv{} of a \ns{} standard module are related by
\begin{equation}
	\sccdim{p}{r}{s} = \slcdim{r}{s} -\frac{t}{4} j^2,
\end{equation}
with a similar formula for Ramond standard modules. As $j$ varies continuously, these relations describe a parabola for each family of standard modules sharing the same $[i]$- and $(r,s)$-labels (we identify $(r,s)$ with $(u-r,v-s)$ of course).  The parabolae corresponding to the same $[i]$ but different $(r,s)$ therefore do not intersect, hence there are no isomorphisms between the typical $\minmod{u}{v}$-modules.  One can also check that applying spectral flow does not give any new irreducible standards.

Given $i$ and $(r,s)$, the atypical standard modules $\scarelrnp{i}{p}{r,s}$, $p\in\lambda_{r,s}+i+2\ZZ$, correspond to certain isolated points in each parabola.  Their subquotients \eqref{eq:sesN2} exhaust the atypical irreducible $\minmod{u}{v}$-modules that we have constructed through branching rules.  The proof that we have found all the irreducible \hw{} $\minmod{u}{v}$-modules now follows from \cite[Thm.~4.3]{CreSch16} as in the unitary case.  However, there are surely other irreducible (non-weight) $\minmod{u}{v}$-modules.  In particular, we expect to be able to construct examples by decomposing tensor products of $\bcghost$-modules with the Whittaker modules of $\slminmod{u}{v}$ recently constructed in \cite{AdaRea17}.  Unfortunately, the relevance of such non-weight modules for \cft{} is not clear to us and so we shall not dwell upon them.

We instead conclude by asking whether a given atypical standard is \hw{} or not.  This is easily answered, for $\scarelrnp{i}{p}{r,s}$, by comparing the conformal dimensions of its submodule ${\scdisrn{i}{p}{r,s}}$ and its quotient ${\scdisrn{i+2}{p+t}{r,s-1}}$.  Using \eqref{eq:Did}, it turns out that ${\scarelrnp{i}{p}{r,s}}$ is a \hw{} $\minmod{u}{v}$-module if and only if $p\ge\lambda_{r,s}+1$.  Otherwise, ${\scarelrnp{i}{p}{r,s}}$ is the contragredient dual of a \hwm{}.

\section{Characters from residues and spectral flows} \label{sec:char}

In this section, we compute the $N=2$ minimal model (super)characters in both the unitary and the non-unitary cases. The tool that we shall use for these computations is the \emph{residue method}, introduced by Gaberdiel and Eholzer \cite{EhoUni96} and outlined below, which allows one to express \(N=2\) (super)characters as residues of \(\slminmod{u}{v}\otimes\bcghost\) (super)characters. In the unitary case, only a certain subset of (super)characters will be computed in this way, with the remainder being then deduced from spectral flow.  We will show, in the non-unitary case, how a certain ``magic identity'' allows us to compute all the (super)characters as residues.  This identity will also be seen to efficiently recover the unitary results.

\subsection{The residue method}

We define the character and supercharacter of a $\minmod{u}{v}$-module \(\Mod{C}\) by
\begin{equation}
	\fch{\Mod{C}}{z;q} = \traceover{\Mod{C}} z^{J_0} q^{L_0^{\supsc} - \cc/24} \quad \text{and} \quad
	\fsch{\Mod{C}}{z;q} = \traceover{\Mod{C}} (-1)^F z^{J_0} q^{L_0^{\supsc} - \cc/24},
\end{equation}
respectively, where $F \in \End(\Mod{C})$ acts as $0$ on the bosonic subspace and as $1$ on the
fermionic subspace.  Consider the branching rule \eqref{eq:generalbranchingrule} for an
\(\slminmod{u}{v}\)-module \(\Mod{M}_\lambda\) with weight support \(\lambda+2\ZZ\):
\begin{equation}\label{eq:generalbranchingrule2}
  \res{(\Mod{M}_{\lambda} \otimes \bcmod{i})} \cong\ \bigoplus_{\mathclap{p \in\lambda+i+2\ZZ}}\ \fock{p} \otimes \tensor*[^{[i]}]{\Mod{C}}{_p^{\Mod{M}}}.
\end{equation}
The (super)character of $\Mod{M}_{\lambda} \otimes \bcmod{i}$ may then be computed as either an \(\slminmod{u}{v}\otimes\bcghost\)-module or as an \(\fboson\otimes\minmod{u}{v}\)-module. The result must be the same, provided that we make the following identification, itself a consequence of the embedding \eqref{eq:theembedding}:
\begin{equation} \label{trace2}
	w^{h_0}x^{Q_0}q^{L_0^{\supsl}+L_0^{\supbc}} = y^{a_0}z^{J_0}q^{L_0^{\supH}+L_0^{\supsc}}
	= (yz^{1/t})^{h_0} (y^2 z^{-k/t})^{Q_0} q^{L_0^{\supsl}+L_0^{\supbc}}.
\end{equation}
In other words, we identify $w$ with $yz^{1/t}$ and $x$ with $y^2 z^{-k/t}$:
\begin{equation}
	\begin{aligned}
	  \fch{\Mod{M}_{\lambda}}{yz^{1/t};q} \fch{\bcmod{i}}{y^2z^{-k/t};q} &= \sum_{\mathclap{p
	      \in\lambda+i+2\ZZ}}\fch{\fock{p}}{y;q}
	  \fch{\tensor*[^{[i]}]{\Mod{C}}{_p^{\Mod{M}}}}{z;q},\\
	  \fch{\Mod{M}_{\lambda}}{yz^{1/t};q} \fsch{\bcmod{i}}{y^2z^{-k/t};q} &=
	  \sum_{\mathclap{p \in\lambda+i+2\ZZ}}\fch{\fock{p}}{y;q}
	  \fsch{\tensor*[^{[i]}]{\Mod{C}}{_p^{\Mod{M}}}}{z;q}.
	\end{aligned}
\end{equation}
The simple form \eqref{charF} of the free boson characters, in particular the fact that they are proportional to a power of \(y\), then implies the following residue formulae for all $p\in\lambda+i+2\ZZ$:
\begin{subequations} \label{eq:mastercharformula}
	\begin{align}
	  \fch{\tensor*[^{[i]}]{\Mod{C}}{_p^{\Mod{M}}}}{z;q}
	  &=\resid{y=0}{\vphantom{\brac[\Big]{}} y^{-p-1}\eta(q)q^{-p^2/4t}\fch{\Mod{M}_\lambda}{yz^{1/t};q}\fch{\bcmod{i}}{y^2z^{-k/t};q}}, \label{eq:reschar}\\
	  \fsch{\tensor*[^{[i]}]{\Mod{C}}{_p^{\Mod{M}}}}{z;q}
	  &=\resid{y=0}{\vphantom{\brac[\Big]{}} y^{-p-1}\eta(q)q^{-p^2/4t}\fch{\Mod{M}_\lambda}{yz^{1/t};q}\fsch{\bcmod{i}}{y^2z^{-k/t};q}}.\label{eq:resschar}
	\end{align}
\end{subequations}

\subsection{Unitary minimal model characters} \label{subsec:unichar}

In \cite{EhoUni96}, the residue formula \eqref{eq:reschar} was used to compute the characters of the vacuum $\minmod{u}{1}$-modules \(\scirrrnu{0}{0}{1}\), specialised to $z=1$.
In this section, we extend their method to calculate unspecialised
(super)character formulae for certain $\minmod{u}{1}$-modules, namely the \(\scirrrnu{i}{0}{r}\), $r=1,\dots,u-1$.
These are precisely the modules that are tensored with the vacuum Fock space $\fock{0}$ in the branching rules \eqref{eq:ubr}. From \cref{subsec:uniminmod}, we know that each spectral flow orbit contains at least one of these modules and so the (super)characters of the remaining modules may be obtained from our results by spectral flow.

In the course of calculating the residue formulae \eqref{eq:mastercharformula} for
\(\scirrrnu{i}{0}{r}\), we shall use the identity
\begin{equation} \label{eq:ehogab}
	\frac{1}{\prod_{i=1}^{\infty}(1-w^2 q^{i-1})(1-w^{-2}q^i)}=\frac{q^{1/12}}{\eta(q)^2}\sum_{\ell\in\ZZ}\phi_{\ell}(q)w^{2\ell} \qquad \text{($\abs{q} < \abs{w}^2 < 1$),}
\end{equation}
where
\begin{equation}\label{eq:phiexp}
	\phi_{\ell}(q)=\sum_{s=0}^{\infty}(-1)^s q^{\ell s +s(s+1)/2}.
\end{equation}
This was derived\footnote{The formula in \cite{EhoUni96} contains a small typo, which we have fixed here, in the exponent of \(q\) in the first factor.} in \cite{EhoUni96} from an identity given in \cite{KacInf84}.  The proof requires some delicacy with convergence regions and we shall take care to respect these in what follows.

Substituting the identity \eqref{eq:ehogab} into the product form of $\fjth{1}{w^2;q}$, the $\slminmod{u}{1}$ character formulae \eqref{slcharuni} becomes
\begin{align}
  \fch{\slirr{r}}{w;q}
  &=\frac{q^{r^2/4u-1/8}}{w}
  \frac{\sum_{j\in\ZZ}q^{j(uj+r)}\brac*{w^{2uj+r}-w^{-2uj-r}}}
    {\prod_{i=1}^\infty\brac*{1-w^2 q^i}\brac*{1-q^i}\brac*{1-w^{-2}q^{i-1}}}\nonumber\\
  &=-\frac{w q^{-1/12}}{\eta(q)}
  \frac{\sum_{j\in\ZZ}q^{(2uj+r)^2/4u}\brac*{w^{2uj+r}-w^{-2uj-r}}}
    {\prod_{i=1}^\infty\brac*{1-w^2 q^{i-1}}\brac*{1-w^{-2}q^{i}}}\nonumber\\
  &=\frac{w}{\eta(q)^3}
  \sum_{j\in\ZZ}q^{(2uj+r)^2/4u}\brac*{w^{-2uj-r}-w^{2uj+r}}
  \sum_{\ell\in\ZZ}\phi_\ell(q)w^{2\ell}.
\end{align}
Combining this with the \ns{} ghost characters \eqref{chbc}, we find that the residue formula \eqref{eq:mastercharformula} for \(\scirrrnu{0}{0}{r}\), with $r$ odd, now yields
\begin{equation}
  \fch{\scirrrnu{0}{0}{r}}{z;q}=\frac{z^{1/u}}{\eta(q)^3} \sum_{\mathclap{j,\ell,n\in \ZZ}} z^{-n}q^{n^2/2+(2uj+r)^2/4u}
  \resid{y=0}{\brac*{yz^{1/u}}^{-2uj-r+2n+2\ell} - \brac*{yz^{1/u}}^{2uj+r+2n+2\ell}} \phi_\ell(q).
\end{equation}
Evaluating the residue then sets \(n=-\frac{1}{2}(1-r)+uj-\ell\) in the first summand and \(n=-\frac{1}{2}(1+r)-uj-\ell\) in the second. The result is thus
\begin{multline}
  \fch{\scirrrnu{0}{0}{r}}{z;q}=\frac{1}{\eta(q)^3}
  \sum_{\mathclap{j,l\in \ZZ}} q^{(2uj+r)^2/4u} \left[ \sum_{s=0}^\infty (-1)^s z^{\ell-uj+(1-r)/2}q^{\ell s+s(s+1)/2+(\ell-uj+(1-r)/2)^2/2} \right. \\
  \left. - \sum_{s=0}^\infty (-1)^s z^{\ell+uj+(1+r)/2}q^{\ell s+s(s+1)/2+(\ell+uj+(1+r)/2)^2/2} \right],
\end{multline}
where we have also substituted the series expansion \eqref{eq:phiexp}.  The exponents of \(z\) and \(q\) in the brackets simplify greatly upon replacing \(\ell\) by \(\ell-s+uj-\frac{1}{2}(1-r)\) in the first summand and by \(\ell-s-uj-\frac{1}{2}(1+r)\) in the second:
\begin{align}\label{eq:nonconvgeosums}
  \fch{\scirrrnu{0}{0}{r}}{z;q}
  &= \frac{1}{\eta(q)^3} \sum_{\mathclap{j,l\in \ZZ}} q^{(2uj+r)^2/4u} \sqbrac*{\sum_{s=0}^\infty (-1)^s z^{\ell-s} q^{\ell^2/2+s(2uj+r)/2} - \sum_{s=0}^\infty (-1)^s z^{\ell-s} q^{\ell^2/2-s(2uj+r)/2}}\nonumber\\
  &=\frac{\fjth{3}{z;q}}{\eta(q)^3} \sum_{j\in\ZZ} q^{(2uj+r)^2/4u}
  \sqbrac*{\sum_{s=0}^\infty \brac*{-z^{-1}q^{(2uj+r)/2}}^s - \sum_{s=0}^\infty \brac*{-z^{-1}q^{-(2uj+r)/2}}^s}.
\end{align}

We have not combined the two sums over $s$ into one, nor have we explicitly summed these geometric series.  This is because their regions of convergence are $j$-dependent and there is no region in which all these geometric series converge simultaneously.
We instead proceed by recalling the product form
\begin{equation}
	\fjth{3}{z;q} = \prod_{i=1}^{\infty} (1+zq^{i-1/2}) (1-q^i) (1+z^{-1}q^{i-1/2})
\end{equation}
and noting the following formal power series identities:
\begin{equation}
	(1-x)\sum_{s=0}^\infty x^s
	= 1,\qquad
	(1-x)\sum_{s=0}^\infty x^{-s}
	= -x,
	\label{eq:geomserid}
\end{equation}
Indeed, $\fjth{3}{z;q}$ will have a factor $(1+z^{-1}q^{(2uj+r)/2})$ if $uj+\frac{r+1}{2} \in \ZZ_{>0}$, that is if $j \in \ZZ_{\ge 0}$, so for these $j$, we may take $x=-z^{-1}q^{(2uj+r)/2}$ to obtain
\begin{equation}
	(1+z^{-1}q^{(2uj+r)/2}) \sum_{s=0}^\infty \brac*{-z^{-1}q^{(2uj+r)/2}}^s = 1.
\end{equation}
Similarly, when $j \in \ZZ_{\ge 0}$, $(1+zq^{(2uj+r)/2})$ is a factor of $\fjth{3}{z;q}$ so putting $x=-zq^{(2uj+r)/2}$ results in
\begin{equation}
	(1+zq^{(2uj+r)/2}) \sum_{s=0}^\infty \brac*{-z^{-1}q^{-(2uj+r)/2}}^s = zq^{(2uj+r)/2}.
\end{equation}
Similarly analysing the $j \in \ZZ_{<0}$ terms leads to the following character formula for $r$ odd:
\begin{multline} \label{unichns}
	\fch{\scirrrnu{0}{0}{r}}{z;q} = \fch{\scirrrnu{2}{0}{r}}{z;q}
	= \frac{q^{\sccdimuni{0}{r} - \cc/24 + 1/8}}{\eta(q)^3} \left[ \sum_{j \ge 0}\ \brac*{\frac{\fjth{3}{z;q}}{1+z^{-1} q^{(2uj+r)/2}} - \frac{z q^{(2uj+r)/2} \, \fjth{3}{z;q}}{1+z q^{(2uj+r)/2}}} q^{j(uj+r)} \right. \\
		\left. + \sum_{j<0}\ \brac*{\frac{z q^{-(2uj+r)/2} \, \fjth{3}{z;q}}{1+z q^{-(2uj+r)/2}} - \frac{\fjth{3}{z;q}}{1+z^{-1} q^{-(2uj+r)/2}}} q^{j(uj+r)} \right].
\end{multline}

This may of course be simplified further.  In particular, we could write \eqref{unichns} in the beguilingly simple form
\begin{equation} \label{eq:unicharsimp}
	\fch{\scirrrnu{0}{0}{r}}{z;q} = \frac{q^{\sccdimuni{0}{r} - \cc/24 + 1/8}}{\eta(q)^3} \sum_{j \in \ZZ}\ \brac*{\frac{\fjth{3}{z;q}}{1+z^{-1} q^{(2uj+r)/2}} - \frac{\fjth{3}{z;q}}{1+z^{-1} q^{-(2uj+r)/2}}} q^{j(uj+r)},
\end{equation}
which matches the result that we would have obtained if we had na\"{\i}vely summed the geometric series in \eqref{eq:nonconvgeosums}.  However, \eqref{unichns} makes manifest the fact that the denominators must be treated as factors of $\fjth{3}{z;q}$.  Because of these cancellations, this character formula is valid for all \(|q|<1\) and \(z \neq 0\).  We therefore conclude that \eqref{eq:unicharsimp} is fine as long as we remember to interpret the terms in parentheses as either being of the form
\begin{equation}
	\frac{\fjth{3}{z;q}}{1+z^{-1} q^{\alpha}} \qquad \text{or} \qquad \frac{z q^{-\alpha} \, \fjth{3}{z;q}}{1+z q^{-\alpha}},
\end{equation}
where the choice is made according as to which denominator is a factor of $\fjth{3}{z;q}$.

To compute the corresponding supercharacter, it suffices to note that taking the supertrace is equivalent to factorising out \(z\) to the power of the charge of the \hwv{} and replacing \(z\) by \(-z\) in what remains. Since the charge of the \hwv{} of \(\scirrrnu{0}{0}{r}\) is $0$, we obtain
\begin{align} \label{unischns}
  \fsch{\scirrrnu{0}{0}{r}}{z;q} &= -\fsch{\scirrrnu{2}{0}{r}}{z;q}=\fch{\scirrrnu{0}{0}{r}}{-z;q}\nonumber\\
  &= \frac{q^{\sccdimuni{0}{r} - \cc/24 + 1/8}}{\eta(q)^3} \sum_{j \in \ZZ}\ \brac*{\frac{\fjth{4}{z;q}}{1-z^{-1} q^{(2uj+r)/2}} - \frac{\fjth{4}{z;q}}{1-z^{-1} q^{-(2uj+r)/2}}} q^{j(uj+r)},
\end{align}
again for $r$ odd and again with the interpretation that the denominators must be turned into factors of $\fjth{4}{z;q}$.  We can repeat these calculations in the Ramond sector for the \(\scirrrnu{i}{0}{r}\), with $i=1,3$ and \(r\) even.  The characters are given by \eqref{unichns}, but with $\jth{3}$ replaced by $\jth{2}$, and the supercharacters are given by \eqref{unischns}, but with $\jth{4}$ replaced by $\ii \jth{1}$.

These equations provide character and supercharacter formulae for the $\minmod{u}{1}$-modules ${\scirrrnu{i}{p}{r}}$ with $p=0$. We recall from \cref{subsec:uniminmod} that these modules are representatives for the spectral flow orbits on the set of all (isomorphism classes of) irreducible modules.  We can therefore use spectral flow to compute the (super)character of every irreducible $\minmod{u}{1}$-module.  The formula relating the characters of a module $\Mod{L}$ and its spectral flows is easily derived:
\begin{align}
  \fch{\sfscsymb^{\ell}(\Mod{L})}{z;q}&=\traceover{\sfscsymb^{\ell}(\Mod{L})}\left[z^{J_0}q^{L^{\supsc}_0-\cc/{24}}
    \right]
  =\traceover{\Mod{L}}\left[z^{\sfscsymb^{-\ell}(J_0)}q^{\sfscsymb^{-\ell}(L^{\supsc}_0)-\cc/{24}}
    \right]\nonumber\\
  &=
  \traceover{\Mod{L}}\left[z^{J_0+\cc\ell/3}q^{L^{\supsc}_0+\ell J_0+\cc\ell^2/6-\cc/{24}}
    \right]
  =z^{\cc\ell/3}q^{\cc\ell^2/6}\fch{\Mod{L}}{zq^\ell;q}.\label{modchar1}
\end{align}
We therefore obtain
\begin{subequations} \label{eq:charsfrel}
	\begin{gather}
		\fch{\scirrrnu{i}{p}{r}}{z;q} = \fch{\sfscsymb^{-p/2}(\scirrrnu{i-p}{0}{r})}{z;q}
		= z^{-p\cc/6}q^{p^2\cc/24}\fch{\scirrrnu{i-p}{0}{r}}{zq^{-p/2};q},
		\intertext{and, similarly,}
		\fsch{\scirrrnu{i}{p}{r}}{z;q} = z^{-p\cc/6}q^{p^2\cc/24}\fsch{\scirrrnu{i-p}{0}{r}}{zq^{-p/2};q}.
	\end{gather}
\end{subequations}

An explicit character formula for the $\scirrrnu{i}{p}{r}$, $p \in r-1+i+2\ZZ$, is thus given by
\begin{equation} \label{eq:genunicharsimp}
	\fch{\scirrrnu{i}{p}{r}}{z;q} = \frac{z^{p/u} q^{\sccdimuni{p}{r} - \cc/24 + 1/8}}{\eta(q)^3} \sum_{j \in \ZZ}\ \brac*{\frac{\fjth{3}{z;q}}{1+z^{-1} q^{(2uj+r+p)/2}} - \frac{\fjth{3}{z;q}}{1+z^{-1} q^{-(2uj+r-p)/2}}} q^{j(uj+r)},
\end{equation}
if $i$ is even, and by the same formula but with $\jth{3}$ replaced by $\jth{2}$, if $i$ is odd.  The formula for the general supercharacter is similar, though a little more complicated, and is left as an exercise.  We emphasise that these formulae must converge for all $z \neq 0$ and $\abs{q}<1$.  Each denominator in the sum should therefore be manipulated, as before, to get a factor of $\fjth{3}{z;q}$ or $\fjth{2}{z;q}$, as appropriate.

We note that these characters (and supercharacters) may also be expressed in terms of the higher-level Appell-Lerch sums of \cite{SemHig05}:
\begin{equation} \label{eq:Appell}
	\AL_n(x,y;q) = \sum_{j \in \ZZ} \frac{x^{nj} q^{nj^2/2}}{1-xyq^j} \qquad \text{($n \in \ZZ_{>0}$).}
\end{equation}
Here, the interpretation of the denominator is again subtle, requiring a geometric series expansion in different regions according as to the sign of $j$.  Explicitly, we have
\begin{equation}
	\AL_n(x,y;q) = \sqbrac[\Big]{\sum_{i,j \ge 0} - \sum_{i,j \le -1}} x^{i+nj} y^i q^{ij + nj^2/2}.
\end{equation}
The characters for $i$ even may therefore be written in the form
\begin{multline} \label{eq:ALunichars}
	\fch{\scirrrnu{i}{p}{r}}{z;q} = z^{p/u} q^{\sccdimuni{p}{r} - \cc/24 + 1/8} \frac{\fjth{3}{z;q}}{\eta(q)^3} \\
	\cdot \sqbrac[\Big]{\AL_2(q^{r/2},-z^{-1} q^{p/2};q^u) - z q^{(r-p)/2} \AL_2(q^{(r+u)/2},-zq^{-(p+u)/2};q^u)}
\end{multline}
and those for $i$ odd follow by replacing $\jth{3}$ by $\jth{2}$.  The corresponding supercharacters are likewise easily found.

One advantage of this reformulation is that the modular properties of the Appell-Lerch sums are known and so can be used to investigate the modularity of these characters.  We shall not do so here, referring instead to the original sources \cite{RavMod87,QiuMod87} and to the more recent treatments \cite{KacInt94,SemHig05,SatMod17,AlfMoc14}.  Another advantage is that it is now straightforward to check that these characters respect the periodicity properties \eqref{eq:uKacSymm}.  This follows from the ``open quasiperiodicity'' property for Appell-Lerch sums given in \cite[Eq.(2.5)]{SemHig05}.

\subsection{Non-unitary minimal model characters}

We now turn to the computation of the (super)characters of the standard modules of the non-unitary $N=2$ minimal models $\minmod{u}{v}$, $v>1$, again by taking residues of $\slminmod{u}{v}\otimes \bcghost$ characters.  While this is straightforward, determining character formulae for the atypical irreducible $\minmod{u}{v}$-modules is much more subtle.  We shall first follow a procedure \cite{CreRel11,CreLog13} in which each atypical irreducible is resolved in terms of atypical standard modules. The character of the former then follow from the Euler-Poincar\'{e} principle, if the resolution converges.  Unfortunately, we shall see that it only does if $k<0$.

As in the unitary case, substituting the character formulae \eqref{charE} for the standard \(\slminmod{u}{v}\)-modules into the residue formulae \eqref{eq:mastercharformula} yields formulae for the \(\minmod{u}{v}\)-(super)characters.  Indeed, these residue formulae are significantly easier to evaluate than those encountered in the unitary case.  This is because the standard \(\slminmod{u}{v}\)-characters contain the algebraic delta function \(\delta(w^2)=\sum_{n\in\ZZ}w^{2n}\) as a factor.  For the typicals, we have
\begin{subequations}\label{eq:n2Echar}
	\begin{align}
		\fch{\screlrn{0}{p}{r,s}}{z;q}
		&= \fch{\screlrn{2}{p}{r,s}}{z;q}
		= \frac{\eta(q)}{q^{p^2/4t}} \resid{y=0}{\vphantom{\brac[\Big]{}} y^{-p-1} \fch{\slrel{p,\slcdim{r}{s}}}{yz^{1/t};q}\fch{\bcmod{0}}{y^2z^{-k/t};q}}\label{eq:n2Echar0}\nonumber\\
		&= \frac{1}{q^{p^2/4t}} \frac{\fchvir{r,s}}{\eta(q)^2} \resid{y=0}{y^{-p-1} \brac*{yz^{1/t}}^{p}\delta\brac*{y^2z^{2/t}}\fjth{3}{y^2z^{-k/t};q}}\nonumber\\
		&= \frac{z^{p/t}}{q^{p^2/4t}} \frac{\fchvir{r,s}}{\eta(q)^2} \resid{y=0}{\vphantom{\brac[\Big]{}} y^{-1} \delta\brac*{y^2z^{2/t}}} \fjth{3}{z^{-1};q}
		=\frac{z^{p/t}}{q^{p^2/4t}} \frac{\fjth{3}{z;q} \fchvir{r,s}}{\eta(q)^2}.
	\end{align}
	and, similarly,
  \begin{align}
    \fsch{\screlrn{0}{p}{r,s}}{z;q} = -\fsch{\screlrn{2}{p}{r,s}}{z;q} &= \frac{z^{p/t}}{q^{p^2/4t}}\frac{\fjth{4}{z;q}\fchvir{r,s}}{\eta(q)^2}, \\
    \fch{\screlrn{1}{p}{r,s}}{z;q} = \fch{\screlrn{3}{p}{r,s}}{z;q} &= \frac{z^{p/t}}{q^{p^2/4t}}\frac{\fjth{2}{z;q}\fchvir{r,s}}{\eta(q)^2},\\
    \fsch{\screlrn{1}{p}{r,s}}{z;q} = -\fsch{\screlrn{3}{p}{r,s}}{z;q} &= \frac{z^{p/t}}{q^{p^2/4t}}\frac{\ii\fjth{1}{z;q}\fchvir{r,s}}{\eta(q)^2}.
  \end{align}
\end{subequations}
These formulae also apply to the atypical standard $\minmod{u}{v}$-modules $\scarelrn{i}{p}{r,s}$, $p \in \lambda_{r,s}+i+2\ZZ$.

The short exact sequences \eqref{eq:sesN2} may be spliced together to form resolutions for the atypical irreducible $\minmod{u}{v}$-modules in terms of these atypical standards.  Alternatively, one may obtain these resolutions from the analogous resolutions for the atypical irreducibles of $\slminmod{u}{v}$ \cite[Prop.~8]{CreMod13} by tensoring with a fixed $\bcghost$-module $\bcmod{i}$, applying the branching rules \eqref{eq:nubr} and \eqref{br:atypst}, and projecting onto a given eigenspace of $a_0$.  For example, either method results in the following resolution for the ${\scirrrn{i}{p}{r,0}}$, \(p\in r-1+i + 2 \ZZ\):
\begin{align} \label{res:L}
	\cdots \lra \scarelrnp{i}{p-(3v-1)t}{r,v-1} \lra \cdots \lra \scarelrnp{i}{p-(2v+2)t}{r,2} &\lra \scarelrnp{i}{p-(2v+1)t}{r,1} \notag \\
	\lra \scarelrnp{i}{p-(2v-1)t}{u-r,v-1} \lra \cdots \lra \scarelrnp{i}{p-(v+2)t}{u-r,2} &\lra \scarelrnp{i}{p-(v+1)t}{u-r,1} \notag \\
	\lra \scarelrnp{i}{p-(v-1)t}{r,v-1}	\lra \cdots \lra \scarelrnp{i}{p-2t}{r,2} &\lra \scarelrnp{i}{p-t}{r,1} \lra \scirrrn{i}{p}{r,0} \lra 0.
\end{align}
Applying Euler-Poincar\'{e} then gives the character of the atypical irreducible $\scirrrn{i}{p}{r,0}$ as an alternating sum of characters of atypical standards.  In particular, we find that
\begin{align} \label{eq:divergent}
	\fch{\scirrrn{0}{p}{r,0}}{z;q}
	&= \sum_{s=1}^{v-1}(-1)^{s-1} \sum_{\ell=0}^{\infty} \brac[\Big]{\fch{\scarelrnp{0}{p-(2v\ell+s)t}{r,s}}{z;q} - \fch{\scarelrnp{0}{p-(2v(\ell+1)-s)t}{u-r,v-s}}{z;q}} \notag \\
	&= \frac{\vartheta_{3}(z;q)}{\eta(q)^2} \sum_{s=1}^{v-1} (-1)^{s-1} \fchvir{r,s} \sum_{\ell=0}^\infty \brac*{\frac{z^{p/t-(2v\ell+s)}}{q^{(p-(2v\ell+s)t)^2/4t}} - \frac{z^{p/t-(2v(\ell+1)-s)}}{q^{(p-(2v(\ell+1)-s)t)^2/4t}}},
\end{align}
where we have substituted the character formula \eqref{eq:n2Echar0}.  The formula for $\scirrrn{1}{p}{r,0}$ may be obtained by replacing $\jth{3}$ by $\jth{2}$, as usual.  Supercharacters also follow straightforwardly.

It is easy to derive similar resolutions for the $\scdisrn{i}{p}{r,s}$, $p \in i+\lambda_{r,s} + 2\ZZ$, and thence arrive at character formulae.  We give the result for $i=0$ for completeness:
\begin{align}
	\fch{\scdisrn{0}{p}{r,s}}{z;q} &=(-1)^{v-1-s}\fch{\scirrrn{0}{p-(v-s)t}{u-r,0}}{z;q}+\sum_{j=1}^{v-s-1}(-1)^{j-1}\fch{\scarelrnp{0}{p-tj}{r,s+j}}{z;q}\notag\\
	&= \frac{\vartheta_3(z;q)}{\eta(q)^2} \left[ \sum_{s'=1}^{v-1} (-1)^{s'+v-s} \fchvir{r,v-s'} \sum_{\ell=0}^\infty \brac*{\frac{z^{p/t-(v-s)-(2v\ell+s')}}{q^{(p-(v-s)t-(2v\ell+s')t)^2/4t}} - \frac{z^{p/t-(v-s)-(2v(\ell+1)-s')}}{q^{(p-(v-s)t-(2v(\ell+1)-s')t)^2/4t}}} \right. \notag\\
	&\mspace{50mu} \left. + \sum_{s'=1}^{v-s-1} (-1)^{s'-1} \fchvir{r,s+s'} \frac{z^{p/t-s'}}{q^{(p-ts')^2/4t}} \right].
\end{align}
We note however that the infinite sums in these formulae do not converge in the required region $0 < \abs{q} < 1$ because $t>0$.  More importantly, it is easy to check that these character formulae do not even converge as formal power series in $z$ (whose coefficients must converge for $0 < \abs{q} < 1$) unless $k<0$.  This unfortunate observation means that the resolutions \eqref{res:L}, as well as their analogues for the other atypical irreducibles, do not converge when $k>0$.  A similar issue was noted recently with the atypical characters of the non-unitary parafermion cosets of \cite{AugMod17}.  The root cause is of course that the resolutions given in \cite{CreMod13} for the atypical $\slminmod{u}{v}$-modules are only convergent when $k<0$.

\subsection{Atypical characters via decomposing meromorphic Jacobi forms} \label{subsec:jacobi}

To circumvent this problem with divergent resolutions for atypical $\minmod{u}{v}$-modules, we reconsider the atypical irreducible characters of the $\SLA{sl}{2}$ minimal models $\slminmod{u}{v}$.  These characters may be analytically continued to meromorphic vector-valued Jacobi forms of weight $0$ and index $k$ \cite{KacMod88}. The decomposition of these forms is rather subtle..  As we have seen, the resolution trick of \cite{CreRel11,CreLog13} fails for $k>0$ and so we need to find another way to solve this problem. In principle, one can answer this question with careful contour integrals as explained in \cite{Dabholkar:2012nd}. Here, we find a much more direct derivation (which also works for $k<0$).

This derivation uses the following ``magic identity'' \cite[Eq.~(A.3)]{CreMod12}, which was itself deduced from \cite[Eq.~(4.8)]{KacInt94}:
\begin{equation} \label{eq:magic}
	\frac{\fjth{1}{ab;q} \eta(q)^3}{\fjth{1}{a;q}\fjth{1}{b;q}} = -\ii \sum_{m \in \ZZ} \frac{a^m}{1 - b q^m} \qquad \text{($\abs{q} < \abs{a} < 1$).}
\end{equation}
This was vital for computing the characters of the atypical standard $\slminmod{u}{v}$-modules (with $v \ge 2$) in \cite{CreMod12,CreMod13}.  We replace $b$ by $-b$ and $-bq^{1/2}$ and simplify, arriving at
\begin{subequations}
	\begin{align}
		\frac{\fjth{2}{ab;q}}{\ii \fjth{1}{a;q}} &= -\frac{\fjth{2}{b;q}}{\eta(q)^3} \sum_{m \in \ZZ} \frac{a^m}{1 + b q^m} & &\text{($\abs{q} < \abs{a} < 1$)} \label{eq:ALidR} \\
		\text{and} \qquad \frac{\fjth{3}{ab;q}}{\ii \fjth{1}{a;q}} &= -\frac{\fjth{3}{b;q}}{\eta(q)^3} \sum_{m \in \ZZ} \frac{a^{m+1/2}}{1 + b q^{m+1/2}} & &\text{($\abs{q} < \abs{a} < 1$)}, \label{eq:ALidNS}
	\end{align}
\end{subequations}
respectively.  As in \cref{subsec:unichar}, the denominators appearing on the \rhs s of these equations should be interpreted as factors of the theta function in the corresponding prefactor.

Consider now the character of the atypical irreducible $\minmod{u}{v}$-module $\scirrrn{0}{p}{r,0}$, $p \in r-1+2\ZZ$.  Combining \eqref{chbc} and \eqref{charL} with the residue formula \eqref{eq:reschar}, we can write this in the form
\begin{align}
	\fch{\scirrrn{0}{p}{r,0}}{z;q}
	&= \residue_{y=0} \sqbrac[\Big]{y^{-p-1} \eta(q) q^{-p^2/4t} \fch{\slirr{r,0}}{yz^{1/t};q} \fch{\bcmod{0}}{y^2z^{-k/t};q}} \notag \\
	&= -q^{\sccdim{p}{r}{0} - \cc/24 + 1/8} \frac{\fjth{3}{z^{-1};q}}{\eta(q)^3} \notag \\
	&\quad \cdot \sum_{j,m \in \ZZ} \residue_{y=0} \sqbrac*{\frac{(yz^{1/t})^{2m+1} z^{-1} q^{m+1/2}}{1 + z^{-1} q^{m+1/2}} y^{-p-1} \brac[\big]{(yz^{1/t})^{2uj+r} - (yz^{1/t})^{-2uj-r}} q^{vj(uj+r)}},
\end{align}
where we have also used \eqref{eq:n2cdimsymm} and \eqref{eq:ALidNS}, the latter with $a=w^2=y^2 z^{2/t}$ and $b=z^{-1}$.  Extracting the residue and simplifying results in
\begin{align} \label{eq:n2Lchar}
	\fch{\scirrrn{0}{p}{r,0}}{z;q}
	&= z^{p/t} q^{\sccdim{p}{r}{0} - \cc/24 + 1/8} \frac{\fjth{3}{z;q}}{\eta(q)^3} \sum_{j \in \ZZ} \sqbrac*{\frac{q^{vj(uj+r)}}{1+z^{-1} q^{(p+r)/2+uj}} - \frac{q^{vj(uj+r)}}{1+z^{-1} q^{(p-r)/2-uj}}} \notag \\
	&= z^{p/t} q^{\sccdim{p}{r}{0} - \cc/24 + 1/8} \frac{\fjth{3}{z;q}}{\eta(q)^3} \notag \\
	&\mspace{100mu} \cdot \sqbrac[\Big]{\AL_{2v}(q^{r/2}, -z^{-1} q^{p/2}; q^u) - z q^{(r-p)/2} \AL_{2v}(q^{(r+t)/2}, -z q^{-(p+t)/2}; q^u)}.
\end{align}
For $p \in r + 2 \ZZ$, we use \eqref{eq:ALidR} instead of \eqref{eq:ALidNS} and arrive at the same formula but with $\jth{3}$ replaced by $\jth{2}$.  Supercharacters are now obtained by replacing $z$ by $-z$ everywhere except in the prefactor $z^{p/t}$.  We note that setting $v=1$, hence $t=u$, in these results recovers the corresponding unitary results \eqref{eq:ALunichars}.  Indeed, the approach described here is easily seen to be equivalent to, though more efficient than, the method of Eholzer and Gaberdiel used in \cref{subsec:unichar}.

We remark that the region of validity, $\abs{q} < \abs{a} < 1$, for the identity \eqref{eq:magic} is a subset of the region of validity, $\abs{q} < \abs{w}^2 < \abs{q}^{-1}$, for the character formula \eqref{charL} of the $\slminmod{u}{v}$-module $\slirr{r,0}$ \cite{CreMod13} (recall that we set $a=w^2$ in the above derivation).  One can however replace $a$ by $aq$ in \eqref{eq:magic} to get a slightly different identity with a slightly different region of validity:
\begin{equation} \label{eq:magic'}
	\frac{\fjth{1}{ab;q} \eta(q)^3}{\fjth{1}{a;q}\fjth{1}{b;q}} = -\ii \sum_{m \in \ZZ} \frac{a^m b q^m}{1 - b q^m} \qquad \text{($1 < \abs{a} < \abs{q}^{-1}$).}
\end{equation}
This is also a subset of the region of validity for the character of $\slirr{r,0}$, so we may repeat the above derivation for the characters of the $\minmod{u}{v}$-modules $\scirrrn{i}{p}{r,0}$ using \eqref{eq:magic'} instead of \eqref{eq:magic}.  The results appear slightly different, for instance $i=0$ and $p \in r-1+2\ZZ$ gives
\begin{multline}
	\fch{\scirrrn{0}{p}{r,0}}{z;q} = z^{p/t} q^{\sccdim{p}{r}{0} - \cc/24 + 1/8} \frac{\fjth{3}{z;q}}{\eta(q)^3} \\
	\cdot \sqbrac[\Big]{\AL_{2v}(q^{r/2}, z q^{-p/2}; q^u) - z^{-1} q^{(r+p)/2} \AL_{2v}(q^{(r+t)/2}, -z^{-1} q^{(p-t)/2}; q^u)}
\end{multline}
instead of \eqref{eq:n2Lchar}.  Comparing, we see that the two character formulae are related by simultaneously swapping $z$ with $z^{-1}$ and $p$ with $-p$.  This of course reflects the fact that $\scirrrn{0}{-p}{r,0}$ is the conjugate of $\scirrrn{0}{p}{r,0}$, see \eqref{eq:cosetauts}.

The ``second magic identity'' \eqref{eq:magic'} is required for determining the characters of the remaining atypical irreducible $\minmod{u}{v}$-modules $\scdisrn{i}{p}{r,0}$ because the character formula \eqref{charD} for the $\slminmod{u}{v}$-atypicals $\sldis{r,s}$ is only valid for $1 < \abs{w}^2 < \abs{q}^{-1}$ \cite{CreMod13}.  Using the same method as before, we arrive at
\begin{multline} \label{eq:n2Dchar}
	\fch{\scdisrn{0}{p}{r,s}}{z;q} = z^{p/t} q^{\sccdim{p}{r}{s} - \cc/24 + 1/8} \frac{\fjth{3}{z;q}}{\eta(q)^3} \\
	\cdot \sqbrac[\Big]{\AL_{2v}(q^{(r-ts)/2}, -z q^{-p/2}; q^u) - q^{rs} \AL_{2v}(q^{-(r+ts)/2}, -z q^{-p/2}; q^u)}
\end{multline}
and, as before, the character of $\scdisrn{1}{p}{r,s}$ is obtained by replacing $\jth{3}$ by $\jth{2}$.  Supercharacters follow as usual.  Note that the denominators implicit in the definition \eqref{eq:Appell} of the Appell-Lerch sums can always be interpreted as a factor of $\fjth{3}{z;q}$.  It follows that these (super)character formulae converge for all $0 < \abs{q} < 1$ and $z \neq 0$, as expected.

\section{Fusion rules} \label{sec:fusion}

One common means of computing the fusion rules of a \cft{} involves finding the characters of its modules and substituting the modular $S$-matrix entries of the characters into the Verlinde formula for fusion coefficients.  This has been proven to work for rational theories \cite{HuaVer05} and seems to also work well for certain logarithmic theories including the $\SLA{sl}{2}$ minimal models $\slminmod{u}{v}$ \cite{CreMod13}.  We therefore expect that this method will also work for the $N=2$ minimal models $\minmod{u}{v}$.  However, this ignores the coset construction technology that we have been exploiting, in particular the branching rules that identify the result of restricting a given $\slminmod{u}{v} \otimes \bcghost$-module to a module over $\fboson \otimes \minmod{u}{v}$.  We will therefore eschew modular methods and compute the fusion rules directly using the fact that the ``inverse'' procedure, known as induction, preserves fusion \cite{RidVer14,CreTen17}.

\subsection{Induction}\label{subsec:induction}

We will start by introducing the induction functor $\ind{}$ that will be used for the calculation of the fusion rules.  Suppose that $\VOA{V}$ is a \svoa{} with subalgebra $\VOA{U}$.  In the application at hand, we will take $\VOA{U} = \fboson\otimes\minmod{u}{v}$ and $\VOA{V} = \slminmod{u}{v}\otimes\bcghost$.  It follows that the restriction $\res{\Mod{B}}$ of a $\VOA{V}$-module $\Mod{B}$ decomposes into a direct sum of $\VOA{U}$-modules.  Contrarily, the \emph{induction} of a $\VOA{U}$-module $\Mod{S}$ is the $\VOA{V}$-module $\ind{\Mod{S}}$ defined by
\begin{equation} \label{eq:inddef}
	\ind{\Mod{S}} = \VOA{V} \fuse \Mod{S},
\end{equation}
where $\fuse$ denotes the fusion product of $\VOA{U}$-modules.  The \rhs{} is indeed a $\VOA{V}$-module because this \svoa{} acts on the first factor of the product (itself in fact).  If we restrict the induced module back to an $\fboson\otimes\minmod{u}{v}$-module again, then the result is
\begin{equation} \label{eq:resind}
	\res{\ind{\Mod{S}}} = \res{\brac*{\VOA{V} \fuse \Mod{S}}} = \res{\VOA{V}} \fuse \Mod{S}.
\end{equation}
In our application, it follows from the first branching rule in \eqref{eq:nubr} that the \rhs{} becomes the direct sum of the fusion products of $\Mod{S}$ with the $\fock{p} \otimes \scirrrn{0}{p}{1,0}$, where $p \in 2\ZZ$.

Let $\bfuse$ denote the fusion product of the $\VOA{V}$-modules.  When we say that induction preserves fusion, we mean that it satisfies
\begin{equation} \label{fusion3}
	\ind{\brac*{\Mod{S}_1 \fuse \Mod{S}_2}} \cong \ind{\Mod{S}_1} \bfuse \ind{\Mod{S}_2},
\end{equation}
for any given $\Mod{U}$-modules $\Mod{S}_1$ and $\Mod{S}_2$ \cite[Thm.~1.4]{CreTen17}.  If we know the inductions of these $\Mod{U}$-modules and the fusion products of the $\Mod{V}$-modules, then it follows that we can determine the inductions of the fusion products of the $\Mod{U}$-modules.  This would not generally suffice to compute the fusion products of $\Mod{U}$ themselves, but the Heisenberg \voa{} is a tensor factor of $\VOA{U}$ in our intended application.  As we shall see, the simplicity of the fusion products \eqref{eq:frH} of its Fock spaces is the key to efficiently computing the fusion products of the other tensor factor $\minmod{u}{v}$.

\subsection{Unitary $N=2$ minimal model fusion rules}\label{subsec:unifusion}

Recall the branching rule \eqref{eq:ubr} of the unitary minimal model $\minmod{u}{1}$.  To compute fusion rules, we shall first need to identify the inductions of the irreducible $\fboson\otimes\minmod{u}{v}$-modules $\fock{p} \otimes \scirrrnu{i}{p}{r}$.  This follows straightforwardly from \eqref{eq:resind} and the fact that the restriction of an irreducible $\slminmod{u}{v}\otimes\bcghost$-module determines it up to isomorphism (including parity):
\begin{align}
	\res{\ind{\brac*{\fock{p} \otimes \scirrrnu{i}{p}{r}}}}
	&\cong \bigoplus_{p' \in 2\ZZ} \brac*{\fock{p} \otimes \scirrrnu{i}{p}{r}} \fuse \brac*{\fock{p'} \otimes \scirrrnu{0}{p'}{1}}
	\cong \bigoplus_{p' \in 2\ZZ} \brac*{\fock{p} \fuse \fock{p'}} \otimes \brac*{\scirrrnu{i}{p}{r} \fuse \scirrrnu{0}{p'}{1}} \notag \\
	&\cong \bigoplus_{p' \in 2\ZZ} \fock{p+p'} \otimes \scirrrnu{i}{p+p'}{r}
	\cong \bigoplus_{\mathclap{p' \in p+2\ZZ}}\ \fock{p'} \otimes \scirrrnu{i}{p'}{r}
	\cong \res{\brac*{\slirr{r}\otimes \bcmod{i}}}.
\end{align}
Here, we note that $p = i+r-1 \bmod{2}$.  The Fock space fusion rules were given in \eqref{eq:frH}, while those involving the $\scirrrnu{0}{p'}{1}$ were evaluated using \cite[Prop.~3.7]{CreSch16}.  It follows that
\begin{equation} \label{eq:indid}
	\ind{\brac*{\fock{p} \otimes \scirrrnu{i}{p}{r}}} \cong \slirr{r}\otimes \bcmod{i},
\end{equation}
for all $r = 1, \dots, u-1$, $i=0,\dots,3$ and $p \in i+r-1+2\ZZ$.

To determine the fusion product of the irreducibles $\scirrrnu{i}{p}{r}$ and $\scirrrnu{i'}{p'}{r'}$, we tensor each with an appropriate Fock space so that the fusion product is
\begin{equation}
	\brac*{\fock{p} \otimes \scirrrnu{i}{p}{r}} \fuse \brac*{\fock{p'} \otimes \scirrrnu{i'}{p'}{r'}}
	\cong \fock{p+p'} \otimes \brac*{\scirrrnu{i}{p}{r} \fuse \scirrrnu{i'}{p'}{r'}}.
\end{equation}
Inducing, and applying \eqref{fusion3}, this becomes
\begin{equation}
	\ind{\brac*{\fock{p} \otimes \scirrrnu{i}{p}{r}}} \bfuse \ind{\brac*{\fock{p'} \otimes \scirrrnu{i'}{p'}{r'}}}
	\cong \brac*{\slirr{r}\otimes \bcmod{i}} \bfuse \brac*{\slirr{r'}\otimes \bcmod{i'}}
	\cong \bigoplus_{r''=1}^{u-1} \vircoe{u}{r''}{r,r'} \slirr{r''}\otimes \bcmod{i+i'},
\end{equation}
where we have used \eqref{eq:indid} and the fusion rules \eqref{eq:bcfus} and \eqref{eq:sl2fusuni}.  We now restrict back to an $\fboson\otimes\minmod{u}{v}$-module.  Using \eqref{eq:resind}, the \lhs{} becomes
\begin{align}
	\res{\ind{\brac*{\fock{p+p'} \otimes \brac*{\scirrrnu{i}{p}{r} \fuse \scirrrnu{i'}{p'}{r'}}}}}
	&\cong \bigoplus_{p'' \in 2\ZZ} \fock{p+p'+p''} \otimes \brac*{\scirrrnu{i}{p}{r} \fuse \scirrrnu{i'}{p'+p''}{r'}} \notag \\
	\cong \bigoplus_{\mathclap{p'' \in p+p'+2\ZZ}}\ \fock{p''} \otimes \brac*{\scirrrnu{i}{p}{r} \fuse \scirrrnu{i'}{p''-p}{r'}}
	&\cong \bigoplus_{\mathclap{p'' \in i+i'+r+r'+2\ZZ}}\ \fock{p''} \otimes \brac*{\scirrrnu{i}{p}{r} \fuse \scirrrnu{i'}{p''-p}{r'}},
\end{align}
since $p \in i+r-1+2\ZZ$ and $p' \in i'+r'-1+2\ZZ$, while the \rhs{} becomes
\begin{align}
	\bigoplus_{r''=1}^{u-1} \vircoe{u}{r''}{r,r'} \res{\brac*{\slirr{r''}\otimes \bcmod{i+i'}}}
	&\cong \bigoplus_{r''=1}^{u-1} \vircoe{u}{r''}{r,r'} \bigoplus_{\mathclap{p'' \in i+i'+r''-1+2\ZZ}}\ \fock{p''} \otimes \scirrrnu{i+i'}{p''}{r''} \notag \\
	&\cong \bigoplus_{\mathclap{p'' \in i+i'+r+r'+2\ZZ}}\ \fock{p''} \otimes \sqbrac*{\bigoplus_{r''=1}^{u-1} \vircoe{u}{r''}{r,r'} \scirrrnu{i+i'}{p''}{r''}},
\end{align}
since the fusion coefficient $\vircoe{u}{r''}{r,r'}$ vanishes unless $r'' = r+r' \bmod{2}$, by \eqref{eq:slfuscoeffs}.  Projecting onto the $a_0$-eigenspace of eigenvalue $p+p'$ therefore gives the fusion rules of the unitary models $\minmod{u}{1}$:
\begin{equation} \label{fr:unitary}
	\scirrrnu{i}{p}{r} \fuse \scirrrnu{i'}{p'}{r'} \cong \bigoplus_{r''=1}^{u-1} \vircoe{u}{r''}{r,r'} \scirrrnu{i+i'}{p+p'}{r''}
\end{equation}

As discussed in \cref{subsec:uniminmod}, there are isomorphisms \eqref{eq:uKacSymm} among the $\minmod{u}{1}$-modules $\scirrrnu{i}{p}{r}$ that appear on the \rhs{}.  One can also rewrite these remarkably simple fusion rules using the dictionary \eqref{eq:unitarydictionary} to translate the notation for the modules into the ``native'' $N=2$ notation (wherein modules are parametrised by the charge $j$ and conformal dimension $\Delta$ of the \hwv{}).  This would have the effect of unnecessarily complicating the fusion rules and so we shall leave such a translation as an exercise for readers who need it for applications.

\subsection{Non-unitary $N=2$ minimal model fusion rules}

The fusion rules for the non-unitary minimal models $\minmod{u}{v}$, $v>1$, may be computed, in principle, using the same technique.  Unfortunately, we do not know the fusion rules of the corresponding $\SLA{sl}{2}$ minimal models in general, but only their Grothendieck counterparts (reported in \cref{app:frsl2}).  These were obtained in \cite{CreMod13} from a (conjectural) version of the Verlinde formula \cite{CreLog13,RidVer14}.  This means that we have the images of the fusion products in the ring (the Grothendieck fusion ring) obtained from the genuine fusion ring by identifying each indecomposable with the sum of its composition factors.  The image of a module $\Mod{M}$ in the Grothendieck fusion ring will be denoted by $\Gr{\Mod{M}}$ and the Grothendieck fusion product by $\Grfuse$.

The induction-restriction method that we have detailed in the unitary cases therefore only allows us to compute the Grothendieck fusion rules of $\minmod{u}{v}$, these being the explicit decomposition of each
\begin{equation}
	\Gr{\Mod{M}} \Grfuse \Gr{\Mod{N}} \equiv \Gr{\Mod{M} \fuse \Mod{N}}
\end{equation}
into sums of images of irreducibles.  Aside from this, the only new feature that appears in the computations, as compared with the unitary computations detailed above, is the need to use branching rules for spectral flows of $\slminmod{u}{v}$-modules.  This follows easily from \eqref{eq:cosetauts} as in \cref{sec:branch}:
\begin{align}
	\res{\brac*{\sfsl{\ell}{\Mod{M}} \otimes \bcmod{i}}}
	&\cong \res{\brac*{\sfsl{\ell}{\Mod{M}} \otimes \sfbc{\ell}{\bcmod{i-2\ell}}}}
	\cong \bigoplus_{\mathclap{p \in i-2\ell+L}}\ \sffb{\ell t}{\fock{p}} \otimes {}^{[i-2\ell]} \Mod{C}^{\Mod{M}}_p \notag \\
	&\cong \bigoplus_{\mathclap{p \in i-2\ell+L}}\ \fock{p+\ell t} \otimes {}^{[i-2\ell]} \Mod{C}^{\Mod{M}}_p
	\cong \bigoplus_{\mathclap{p \in i+\ell k+L}}\ \fock{p} \otimes {}^{[i-2\ell]} \Mod{C}^{\Mod{M}}_{p-\ell t}.
\end{align}
Here, $L$ denotes the set of $\SLA{sl}{2}$-weights of the $\slminmod{u}{v}$-module $\Mod{M}$.  With this in hand, the Grothendieck fusion rules are as follows:
\begin{subequations}
	\begin{align}
		\Gr{\scirrrn{i}{p}{r,0}} \Grfuse \Gr{\scirrrn{i'}{p'}{r',0}}
		&= \sum_{r''} \vircoe{u}{r''}{r,r'} \Gr{\scirrrn{i+i'}{p+p'}{r'',0}}, \label{LLGrfus} \\
		\Gr{\scirrrn{i}{p}{r,0}} \Grfuse \Gr{\screlrn{i'}{p'}{r',s'}}
		&= \sum_{r''} \vircoe{u}{r''}{r,r'} \Gr{\screlrn{i+i'}{p+p'}{r'',s'}}, \\
		\Gr{\scirrrn{i}{p}{r,0}} \Grfuse \Gr{\scdisrn{i'}{p'}{r',s'}}
		&= \sum_{r''} \vircoe{u}{r''}{r,r'} \Gr{\scdisrn{i+i'}{p+p'}{r'',s'}}, \\
		\Gr{\screlrn{i}{p}{r,s}} \Grfuse \Gr{\screlrn{i'}{p'}{r',s'}}
		&= \sum_{r'',s''} \vircoe{u}{r''}{r,r'} \vircoe{v}{s''}{s,s'} \brac*{\Gr{\screlrn{i+i'-2}{p+p'-t}{r'',s''}} + \Gr{\screlrn{i+i'+2}{p+p'+t}{r'',s''}}} \notag \\
		&\mspace{100mu} + \sum_{r'',s''} \vircoe{u}{r''}{r,r'} \brac*{\vircoe{v}{s''}{s,s'-1} + \vircoe{v}{s''}{s,s'+1}} \Gr{\screlrn{i+i'}{p+p'}{r'',s''}}, \label{EEGrfus} \\
		\Gr{\screlrn{i}{p}{r,s}} \Grfuse \Gr{\scdisrn{i'}{p'}{r',s'}}
		&= \sum_{r'',s''} \vircoe{u}{r''}{r,r'} \vircoe{v}{s''}{s,s'+1} \Gr{\screlrn{i+i'}{p+p'}{r'',s''}} \notag \\
		&\mspace{100mu} + \sum_{r'',s''} \vircoe{u}{r''}{r,r'} \vircoe{v}{s''}{s,s'} \Gr{\screlrn{i+i'-2}{p+p'-t}{r'',s''}}, \label{EDGrfus} \\
		\Gr{\scdisrn{i}{p}{r,s}} \Grfuse \Gr{\scdisrn{i'}{p'}{r',s'}} &=
		\begin{dcases*}
			\sum_{r'',s''} \vircoe{u}{r''}{r,r'} \vircoe{v}{s''}{s,s'} \Gr{\screlrn{i+i'-2}{p+p'-t}{r'',s''}} \\
			\mspace{80mu} + \sum_{r''} \vircoe{u}{r''}{r,r'} \Gr{\scdisrn{i+i'}{p+p'}{r'',s+s'}}, & if $s+s'<v$, \\
			\sum_{r'',s''} \vircoe{u}{r''}{r,r'} \vircoe{v}{s''}{s+1,s'+1} \Gr{\screlrn{i+i'-2}{p+p'-t}{r'',s''}} \\
			\mspace{80mu} +\sum_{r''} \vircoe{u}{r''}{r,r'} \Gr{\scdisrn{i+i'-2}{p+p'-t}{u-r'',s+s'-v+1}}, & if $s+s'\ge v$.
		\end{dcases*} \label{DDGrfus}
	\end{align}
\end{subequations}
Here, sums over $r''$ run from $1$ to $u-1$ and sums over $s''$ run from $1$ to $v-1$.

The Grothendieck fusion rules of the $\scirrrn{i}{p}{r,0}$ in fact lift to genuine fusion rules for $\minmod{u}{v}$.  This follows from the fact that the same is true for $\slminmod{u}{v}$, see \eqref{eq:frsl2Lx}, and the fact that Heisenberg cosets preserve module structures \cite[Thm.~3.8]{CreSch16}.  In particular, this gives the following $\minmod{u}{v}$ fusion rules:
\begin{align}
	\scirrrn{i}{p}{r,0} \fuse \scirrrn{i'}{p'}{r',0} &\cong \bigoplus_{r''=1}^{u-1} \vircoe{u}{r''}{r,r'} \scirrrn{i+i'}{p+p'}{r'',0}, \label{LLfus} \\
	\scirrrn{i}{p}{r,0} \fuse \screlrn{i'}{p'}{r',s'} &\cong \bigoplus_{r''=1}^{u-1} \vircoe{u}{r''}{r,r'} \screlrn{i+i'}{p+p'}{r'',s'}, \label{LEfus} \\
	\scirrrn{i}{p}{r,0} \fuse \scdisrn{i'}{p'}{r',s'} &\cong \bigoplus_{r''=1}^{u-1} \vircoe{u}{r''}{r,r'} \scdisrn{i+i'}{p+p'}{r'',s'}.
\end{align}
We remark that if we also assume a vertex tensor category structure on a category containing the $\scirrrn{i}{p}{r,0}$, then the results of \cite{CreSch16,CreBra17} yield a rigorous proof of \eqref{LLfus}, independent of the conjectural standard Verlinde formula.  Comparing this fusion rule with \eqref{fr:unitary}, we see that we have an embedding of the fusion ring of $\minmod{u}{1}$ in that of $\minmod{u}{v}$.  The analogous statement for the $\SLA{sl}{2}$ minimal models was observed in \cite{CreMod13} and for minimal models of simply-laced Lie algebras in general in \cite{CreAdm18}.

Identifying the remaining $\minmod{u}{v}$ fusion rules is more challenging because they are expected to involve reducible but indecomposable modules in general.  Using the conjectural description of the ``staggered'' $\slminmod{u}{v}$-modules reported in \cref{app:frsl2}, we can combine the branching rules \eqref{eq:nubr} with \cite[Thm.~3.8]{CreSch16} to deduce conjectural descriptions of similar $\minmod{u}{v}$-modules which we shall denote by $\scproj{i}{p}{r,s}$, $p \in i + \lambda_{r,s} + 2\ZZ$.  These staggered $\minmod{u}{v}$-modules are reducible but indecomposable, with four composition factors each.  This is summarised in the following Loewy diagrams (we refer to \cite[App.~A.4]{CreLog13} for an accessible review of this concept):
\begin{equation} \label{eq:loewyN=2}
	\begin{tikzpicture}[->,thick,>=latex,baseline=(c.base)]
		\node (top) at (0,2) {$\scdisrn{i}{p}{r,s}$};
		\node (left) at (-2,0) {$\scdisrn{i+2}{p+t}{r,s-1}$};
		\node (right) at (2,0) {$\scdisrn{i-2}{p-t}{r,s+1}$};
		\node (bottom) at (0,-2) {$\scdisrn{i}{p}{r,s}$};
		\draw (top) -- (left);
		\draw (top) -- (right);
		\draw (left) -- (bottom);
		\draw (right) -- (bottom);
		\node[nom] (c) at (0,0) {$\scproj{i}{p}{r,s}$};
	\end{tikzpicture}
	\qquad \text{($s=0,1,\dots,v-1$).}
\end{equation}
To ensure that these Loewy diagrams uniformly describe all the staggered $\minmod{u}{v}$-modules, we have adopted some convenient notation, namely
\begin{equation}
	\scdisrn{i}{p}{r,-1} = \scdisrn{i+2}{p+t}{u-r,v-2}, \qquad
	\scdisrn{i}{p}{r,0} = \scirrrn{i}{p}{r,0}, \qquad \text{and} \qquad
	\scdisrn{i}{p}{r,v} = \scdisrn{i-2}{p-t}{u-r,1}.
\end{equation}
We conjecture that these staggered modules are \emph{projective} in the category of all $\minmod{u}{v}$-modules that belong to a thick version of category $\categ{O}$, for the \ns{} or Ramond $N=2$ Lie superalgebra as appropriate, in which one admits extensions on which $L^{\supsc}_0$ acts with finite-rank Jordan blocks ($J_0$ is still required to act semisimply).

We conclude by illustrating how these staggered $\minmod{u}{v}$-modules are expected to arise in fusion.  Using the same methodology as detailed in \cref{subsec:unifusion}, conjectures for the fusion rules of $\slminmod{u}{v}$ yield (conjectural) $\minmod{u}{v}$ fusion rules.  For example, the conjectural fusion rule \eqref{fr:ExEsl2} (originally made in \cite{CreCos18}) gives
\begin{multline} \label{EEfus}
	\screlrn{i}{p}{1,1} \fuse \screlrn{i'}{p'}{r,s} \\ \cong
	\scalebox{0.9}{$
		\begin{cases*}
			\scproj{i+i'}{p+p'}{r,s-1} \oplus \screlrn{i+i'+2}{p+p'+t}{r,s} \oplus \screlrn{i+i'}{p+p'}{r,s+1}, & if $p+p'-i-i' \in \lambda_{r,s-1} + 2\ZZ$, \\
			\scproj{i+i'}{p+p'}{u-r,v-s-1} \oplus \screlrn{i+i'+2}{p+p'+t}{r,s} \oplus \screlrn{i+i'}{p+p'}{r,s-1}, & if $p+p'-i-i' \in \lambda_{u-r,v-s-1} + 2\ZZ$, \\
			\scproj{i+i'+2}{p+p'+t}{r,s} \oplus \screlrn{i+i'-2}{p+p'-t}{r,s} \oplus \screlrn{i+i'}{p+p'}{r,s-1}, & if $p+p'-i-i' \in \lambda_{r,s+1} + 2 \ZZ$, \\
			\scproj{i+i'+2}{p+p'+t}{u-r,v-s} \oplus \screlrn{i+i'-2}{p+p'-t}{r,s} \oplus \screlrn{i+i'}{p+p'}{r,s+1}, & if $p+p'-i-i' \in \lambda_{u-r,v-s+1} + 2\ZZ$, \\
			\screlrn{i+i'-2}{p+p'-t}{r,s} \oplus \screlrn{i+i'+2}{p+p'+t}{r,s} \oplus \screlrn{i+i'}{p+p'}{r,s-1} \oplus \screlrn{i+i'}{p+p'}{r,s+1}, & otherwise,
		\end{cases*}
	$}
\end{multline}
providing that $2 \le s \le v-2$.  When $s=1$ or $v-1$, we must remove those $\screlrn{i''}{p''}{r,s'}$ with $s'=0$ or $v$ from the \rhs{}.  It may also happen that some of the conditions on $p+p'-i-i'$ coincide, in which case we must remove any direct summands that do not appear in each of the corresponding \rhss{}.

By combining the fusion rules \eqref{LEfus} and \eqref{EEfus}, one can obtain conjectures for all the fusion rules among the typical $\minmod{u}{v}$-modules $\screlrn{i}{p}{r,s}$.  It is also straightforward to deduce conjectures for the remaining fusion rules, meaning those corresponding to the Grothendieck fusion rules \eqref{EDGrfus} and \eqref{DDGrfus}.  We shall not do so here and instead refer the interested reader to \cite{Tian}.

\appendix

\section{Grothendieck fusion rules for the $\SLA{sl}{2}$ minimal models} \label{app:frsl2}

The Grothendieck fusion rules of the non-unitary $\SLA{sl}{2}$ minimal models $\slminmod{u}{v}$, $v>1$, were computed in \cite{CreMod13} using the (conjectural) standard Verlinde formula proposed in \cite{CreLog13,RidVer14}.  We collect the results here for convenience:
\begin{subequations} \label{eq:grfrsl2}
	\begin{align}
		\Gr{\slirr{r,0}} \Grfuse \Gr{\slirr{r',0}}
		&= \sum_{r''} \vircoe{u}{r''}{r,r'} \Gr{\slirr{r'',0}}, \label{grfr:LxL} \\
		\Gr{\slirr{r,0}} \Grfuse \Gr{\slrel{\lambda',\slcdim{r'}{s'}}}
		&= \sum_{r''} \vircoe{u}{r''}{r,r'} \Gr{\slrel{\lambda'+r-1,\slcdim{r''}{s'}}}, \\
		\Gr{\slirr{r,0}} \Grfuse \Gr{\sldis{r',s'}}
		&= \sum_{r''} \vircoe{u}{r''}{r,r'} \Gr{\sldis{r'',s'}}, \\
		\Gr{\slrelp{\lambda,\slcdim{r}{s}}} \Grfuse \Gr{\slrelp{\lambda',\slcdim{r'}{s'}}}
		&= \sum_{r'',s''} \vircoe{u}{r''}{r,r'} \vircoe{v}{s''}{s,s'} \brac*{\Gr{\sfslsymb \brac[\Big]{\slrelp{\lambda+\lambda'-k,\slcdim{r''}{s''}}}} + \Gr{\sfslsymb^{-1} \brac[\Big]{\slrelp{\lambda+\lambda'+k,\slcdim{r''}{s''}}}}} \notag \\
		&\mspace{100mu} + \sum_{r'',s''} \vircoe{u}{r''}{r,r'} \brac*{\vircoe{v}{s''}{s,s'-1} + \vircoe{v}{s''}{s,s'+1}} \Gr{\slrelp{\lambda+\lambda',\slcdim{r''}{s''}}}, \\
		\Gr{\sldis{r,s}} \Grfuse \Gr{\slrelp{\lambda',\slcdim{r'}{s'}}}
		&= \sum_{r'',s''} \vircoe{u}{r''}{r,r'} \vircoe{v}{s''}{s+1,s'} \Gr{\slrelp{\lambda'+\lambda_{r,s},\slcdim{r''}{s''}}} \notag \\
		&\mspace{100mu} + \sum_{r'',s''} \vircoe{u}{r''}{r,r'} \vircoe{v}{s''}{s,s'} \Gr{\sfslsymb \brac[\Big]{\slrelp{\lambda'+\lambda_{r,s+1},\slcdim{r''}{s''}}}}, \\
		\Gr{\sldis{r,s}} \Grfuse \Gr{\sldis{r',s'}} &=
		\begin{dcases*}
			\sum_{r'',s''} \vircoe{u}{r''}{r,r'} \vircoe{v}{s''}{s,s'} \Gr{\sfslsymb \brac[\Big]{\slrelp{\lambda_{r'',s+s'+1},\slcdim{r''}{s''}}}} \\
			\mspace{80mu} + \sum_{r''} \vircoe{u}{r''}{r,r'} \Gr{\sldis{r'',s+s'}}, & if $s+s'<v$, \\
			\sum_{r'',s''} \vircoe{u}{r''}{r,r'} \vircoe{v}{s''}{s+1,s'+1} \Gr{\sfslsymb \brac[\Big]{\slrelp{\lambda_{r'',s+s'+1},\slcdim{r''}{s''}}}} \\
			\mspace{80mu} + \sum_{r''} \vircoe{u}{r''}{r,r'} \Gr{\sfslsymb \brac[\Big]{\sldis{u-r'',s+s'-v+1}}}, & if $s+s'\ge v$.
		\end{dcases*}
	\end{align}
\end{subequations}
In these formulae, $r''$ and $s''$ are summed from $1$ to $u-1$ and from $1$ to $v-1$, respectively.  We mention that these results are consistent with the genuine fusion rules that were computed for $\slminmod{2}{3}$ in \cite{GabFus01} (see \cite{CreMod12} for some corrections) and for $\slminmod{3}{2}$ \cite{RidFus10}.  \cref{grfr:LxL} is also consistent with the genuine fusion rule \eqref{fr:LxL} below, which has recently been proven rigorously for all coprime $u,v \in \ZZ_{\ge 2}$ \cite{CreBra17}.

As was noted in \cite{CreMod13}, the first three Grothendieck fusion rules in \eqref{eq:grfrsl2} actually imply the corresponding genuine fusion rules.  We record these for convenience:
\begin{subequations} \label{eq:frsl2Lx}
	\begin{align}
		\slirr{r,0} \fuse \slirr{r',0} &\cong \bigoplus_{r''=1}^{u-1} \vircoe{u}{r''}{r,r'} \slirr{r'',0}, \label{fr:LxL} \\
		\slirr{r,0} \fuse \slrel{\lambda',\slcdim{r'}{s'}} &\cong \bigoplus_{r''=1}^{u-1} \vircoe{u}{r''}{r,r'} \slrel{\lambda'+r-1,\slcdim{r''}{s'}}, \\
		\slirr{r,0} \fuse \sldis{r',s'} &\cong \bigoplus_{r''=1}^{u-1} \vircoe{u}{r''}{r,r'} \sldis{r'',s'}.
	\end{align}
\end{subequations}
The remaining fusion rules are expected to involve additional reducible, but indecomposable, $\slminmod{u}{v}$-modules with four composition factors each (this has been explicitly verified for $(u,v)=(2,3)$ and $(3,2)$).  They are examples of staggered modules, in the sense of \cite{RidSta09,CreLog13}, possessing a non-diagonalisable action of the Virasoro zero mode $L^{\supsl}_0$.  As such, they are responsible for the logarithmic nature of the corresponding \cfts{}.

It was recently conjectured \cite{CreCos18} that these these staggered modules are projective (in an appropriate category).  To describe them, we introduce the following notation:
\begin{equation} \label{eq:convenientnotation}
	\sldispm{r,-1} = \sldismp{r,1}, \qquad
	\sldis{r,0} \equiv \slirr{r,0} \equiv \sldism{r,0} \qquad \text{and} \qquad
	\sldispm{r,v} = \sfsl{\pm 1}{\sldispm{u-r,1}}.
\end{equation}
The projective whose (unique) irreducible quotient is isomorphic to $\sldispm{r,s}$, for $s=0,1,\dots,v-1$, will be denoted by $\slproj{r,s}^{\pm}$.  We shall sometimes drop the label $\pm$ when $s=0$ in accordance with the second identification of \eqref{eq:convenientnotation}.

The structures of these (conjecturally) projective staggered modules will be characterised in terms of their \emph{Loewy diagrams}; we refer to \cite[App.~A.4]{CreLog13} for an elementary introduction to this concept.  The structural conjecture of \cite{CreCos18} is then that the Loewy diagram of $\slproj{r,s}^{\pm}$ is
\begin{equation} \label{eq:loewysl2}
	\begin{tikzpicture}[->,thick,>=latex,baseline=(c.base)]
		\node (top) at (0,2) {$\sldispm{r,s}$};
		\node (left) at (-2,0) {$\sfsl{\mp1}{\sldispm{r,s-1}}$};
		\node (right) at (2,0) {$\sfsl{\pm1}{\sldispm{r,s+1}}$};
		\node (bottom) at (0,-2) {$\sldispm{r,s}$};
		\draw (top) -- (left);
		\draw (top) -- (right);
		\draw (left) -- (bottom);
		\draw (right) -- (bottom);
		\node[nom] (c) at (0,0) {$\slproj{r,s}^{\pm}$};
	\end{tikzpicture}
	\qquad \text{($s=0,1,\dots,v-1$).}
\end{equation}
(We have taken the opportunity to correct a small typo in the presentation of \cite{CreCos18}.)  The spectral flow images $\sfsl{\ell}{\slproj{r,s}^{\pm}}$ have similar Loewy diagrams that are obtained by applying $\sfslsymb^{\ell}$ to each composition factor in \eqref{eq:loewysl2}.

A subsequent conjecture of \cite{CreCos18} concerns certain genuine fusion rules that involve the staggered projectives introduced above (similar conjectures for the remaining fusion rules will be reported in \cite{Tian}).  Specifically, the fusion rules that generate the typical by typical products were proposed, under the irreducibility assumptions that $\lambda \neq \lambda_{1,1}, \lambda_{u-1,v-1} \bmod{2}$ and $\lambda' \neq \lambda_{r,s}, \lambda_{u-r,v-s} \bmod{2}$.  Then, for all $1 \le r \le u-1$ and $2 \le s \le v-2$ (which requires that $v \ge 4$), the generating fusion rules are conjectured to be
\begin{equation} \label{fr:ExEsl2}
	\slrel{\lambda,\slcdim{1}{1}} \fuse \slrel{\lambda',\slcdim{r}{s}} \cong
	\begin{cases*}
		\slproj{r,s-1}^+ \oplus \sfsl{-1}{\slrel{\lambda+\lambda'+t,\slcdim{r}{s}}} \oplus \slrel{\lambda+\lambda',\slcdim{r}{s+1}}, & if $\lambda+\lambda'=\lambda_{r,s-1}$, \\
		\slproj{u-r,v-s-1}^+ \oplus \sfsl{-1}{\slrel{\lambda+\lambda'+t,\slcdim{r}{s}}} \oplus \slrel{\lambda+\lambda',\slcdim{r}{s-1}}, & if $\lambda+\lambda'=\lambda_{u-r,v-s-1}$, \\
		\slproj{u-r,v-s-1}^- \oplus \sfsl{}{\slrel{\lambda+\lambda'-t,\slcdim{r}{s}}} \oplus \slrel{\lambda+\lambda',\slcdim{r}{s-1}}, & if $\lambda+\lambda'=\lambda_{r,s+1}$, \\
		\slproj{r,s-1}^- \oplus \sfsl{}{\slrel{\lambda+\lambda'-t,\slcdim{r}{s}}} \oplus \slrel{\lambda+\lambda',\slcdim{r}{s+1}}, & if $\lambda+\lambda'=\lambda_{u-r,v-s+1}$, \\
		\sfsl{}{\slrel{\lambda+\lambda'-t,\slcdim{r}{s}}} \oplus \sfsl{-1}{\slrel{\lambda+\lambda'+t,\slcdim{r}{s}}} \oplus \slrel{\lambda+\lambda',\slcdim{r}{s-1}} \oplus \slrel{\lambda+\lambda',\slcdim{r}{s+1}}, & otherwise,
	\end{cases*}
\end{equation}
where $\lambda+\lambda'$ is always understood mod $2$.

When $s=1$ or $s=v-1$, these fusion rules are modified to remove any $\slrel{\lambda'',\slcdim{r}{s'}}$, with $s'=0$ or $v$, and any direct summands that do not appear in all expressions corresponding to the same value of $\lambda+\lambda' \bmod{2}$.  For example, the fusion rule for $s=1$, $v \ge 3$ and $\lambda+\lambda' = \lambda_{r,0} \bmod{2}$ becomes
\begin{equation}
	\slrel{\lambda,\slcdim{1}{1}} \fuse \slrel{\lambda',\slcdim{r}{1}} = \slproj{r,0} \oplus \slrel{\lambda+\lambda',\slcdim{r}{2}},
\end{equation}
because $\lambda_{r,0} = \lambda_{u-r,v}$ and the spectrally flowed summands in the first and fourth cases of \eqref{fr:ExEsl2} are different.  When $v=2$, we would also have to remove the $\slrel{\lambda+\lambda',\slcdim{r}{2}}$ from the \rhs.

\flushleft
\providecommand{\opp}[2]{\textsf{arXiv:\mbox{#2}/#1}}
\providecommand{\pp}[2]{\textsf{arXiv:#1 [\mbox{#2}]}}

\end{document}